\documentclass[twocolumn,trackchanges]{aastex631}

\usepackage{color}
\usepackage{amssymb, amsmath}

\usepackage[utf8]{inputenc}
\usepackage{lipsum}
\usepackage{pifont}
\usepackage{bm}
\usepackage{ulem}
\usepackage{balance}
\usepackage{multirow}
\usepackage{mathtools}
\usepackage{enumitem}
\usepackage{booktabs}
\usepackage{appendix}
\usepackage{listings}
\usepackage{fancyvrb}

\newcommand{\lvalid}{L_{\rm valid}}
\newcommand{\lhole}{L_{\rm hole}}
\newcommand{\lperc}{L_{\rm perc}}
\newcommand{\lstyle}{L_{\rm style}}
\newcommand{\ltv}{L_{\rm tv}}
\newcommand{\ltot}{L_{\rm tot}}
\newcommand{\igt}{\mathbf{I}_{\rm gt}}
\newcommand{\iout}{\mathbf{I}_{\rm out}}
\newcommand{\icomp}{\mathbf{I}_{\rm comp}}

\def\mhmpc{\,{\rm Mpc}/h}
\def\mihmpc{\,h/{\rm Mpc}}

\newcommand\norm[1]{\left\lVert#1\right\rVert}
\def\({\left(}
\def\){\right)}

\graphicspath{{./}{plots/}}
\shorttitle{Deep Inpainting Hydrodynamical Maps}
\shortauthors{Mohammad et al.}

\begin{document}

\title{Inpainting hydrodynamical maps with deep learning}

\correspondingauthor{Faizan G. Mohammad}
\email{faizangohar.mohammad@gmail.com}

\author[0000-0001-9243-7434]{Faizan G. Mohammad}
\affiliation{Waterloo Center for Astrophysics, University of Waterloo, Waterloo, ON N2L 3G1, Canada}
\affiliation{Department of Physics and Astronomy, University of Waterloo, Waterloo, ON N2L 3G1, Canada}

\author[0000-0002-4816-0455]{Francisco Villaescusa-Navarro}
\affiliation{Department of Astrophysical Sciences, Princeton University, Peyton Hall, Princeton NJ 08544-0010, USA}
\affiliation{Center for Computational Astrophysics, Flatiron Institute, 162 5th Avenue, 10010, New York, NY, USA}

\author[0000-0002-3185-1540]{Shy Genel}
\affiliation{Center for Computational Astrophysics, Flatiron Institute, 162 5th Avenue, 10010, New York, NY, USA}
\affiliation{Columbia Astrophysics Laboratory, Columbia University, 550 West 120th Street, New York, NY, 10027, USA}

\author[0000-0001-5769-4945]{Daniel Angl\'es-Alc\'azar}
\affiliation{Department of Physics, University of Connecticut, 196 Auditorium Road, U-3046, Storrs, CT, 06269, USA}
\affiliation{Center for Computational Astrophysics, Flatiron Institute, 162 5th Avenue, 10010, New York, NY, USA}

\author[0000-0001-8593-7692]{Mark Vogelsberger}
\affiliation{Kavli Institute for Astrophysics and Space Research, Department of Physics, MIT, Cambridge, MA 02139, USA}

\begin{abstract}
From 1,000 hydrodynamic simulations of the CAMELS project, each with a different value of the cosmological and astrophysical parameters, we generate 15,000 gas temperature maps. We use a state-of-the-art deep convolutional neural network to recover missing data from those maps. We mimic the missing data by applying regular and irregular binary masks that cover either $15\%$ or $30\%$ of the area of each map. We quantify the reliability of our results using two summary statistics: 1) the distance between the probability density functions (pdf), estimated using the Kolmogorov-Smirnov (KS) test, and 2) the 2D power spectrum. We find an excellent agreement between the model prediction and the unmasked maps when using the power spectrum: better than $1\%$ for $k<20\mihmpc$ for any irregular mask. For regular masks, we observe a systematic offset of $\sim5\%$ when covering $15\%$ of the maps while the results become unreliable when $30\%$ of the data is missing. The observed KS-test p-values favor the null hypothesis that the reconstructed and the ground-truth maps are drawn from the same underlying distribution when irregular masks are used. For regular-shaped masks on the other hand, we find a strong evidence that the two distributions do not match each other. Finally, we use the model, trained on gas temperature maps, to perform inpainting on maps from completely different fields such as gas mass, gas pressure, and electron density and also for gas temperature maps from simulations run with other codes. We find that visually, our model is able to reconstruct the missing pixels from the maps of those fields with great accuracy, although its performance using summary statistics depends strongly on the considered field.
\end{abstract}


%%%%%%%%%%%%%%%%%%%%%%%%%%%%%%%%%%%%%%%%%%%%
\section{Introduction}\label{sec:introduction}					%%%%%%%%%%
%%%%%%%%%%%%%%%%%%%%%%%%%%%%%%%%%%%%%%%%%%%%

Cosmology is in a transformative stage. Nowadays, we know the value of the main cosmological parameters with a relatively high precision. This has allowed us to claim, with high confidence, the existence of a substance that is responsible for the accelerated expansion of the Universe: dark energy. The nature and properties of dark energy remain the biggest mysteries in modern physics. In order to shed light on these and other open questions like the sum of the neutrino masses, the community has spent billions of dollars in surveys like the Dark Energy Spectroscopic Instrument \citep[DESI,][]{DESI_2016}, Euclid \citep{Laureijs_2011}, Prime Focus Spectrograph \citep[PFS,][]{pfs2016}, extended ROentgen Survey with an Imaging Telescope Array \citep[eROSITA,][]{erosita}, Roman Observatory \citep{roman_obs}, Rubin Observatory \citep{LSST}, Square Kilometer Array \citep[SKA,][]{SKA_paper}, and Simons Observatory \citep{simons_obs} whose data may contain the answer to all these fundamental questions. 

The traditional method used to transform the data from  cosmological surveys into constraints is this: 1) the data is compressed into a lower dimension summary statistic, 2) theoretical predictions for that summary statistic are provided as a function of the value of the cosmological parameters, and 3) a likelihood function is evaluated to find the parameter constraints. Currently, there is a large debate on what summary statistics should be employed to extract the maximum information from these surveys \cite[e.g.][]{Quijote, Samushia_2021, Gualdi_2021, Kuruvilla_2021, Bayer_2021,  Banerjee_2019, Changhoon_2019, Uhlemann_2020, Friedrich_2020, Massara_2020, Dai_2020, Allys_2020, Banerjee_2020, Banerjee_2021, Gualdi_2020, Giri_2020, Bella_2020, Changhoon_2020, Valgiannis_2021}. Another possibility is to extract information from the field itself without relying on summary statistics using machine learning methods \citep{Siamak_16, Schmelzle_17, Gupta_18, Ribli_19, Fluri_19, Ntampaka_19, Sultan_2019, Jose_2020, Niall_2020}.

Unfortunately, the data from the cosmic surveys is affected by numerous issues, such as instrument noise. Among these problems, there are some effects that can induce spatial discontinuities in the data. For instance, in the case of galaxy redshift surveys, the presence of stars, fibre collisions, and bad observations, will create masks in the survey geometry \citep{ross12,sdlt13,davide20,faizan20}. Another example is when such masks are created to avoid contamination by systematic effects; e.g. Cosmic Microwave Background (CMB) and 21cm observations may be masked near the galactic plane to avoid the bright foregrounds.

In general, the complicated geometry induced by these masked regions represents a challenge for both the theoretical predictions and the computation of the (optimal) summary statistic. This problem may also get worse when working at the field level with machine learning methods, as one needs to make sure that no information from the mask itself is used by the network.

One potential solution to this problem will be to reconstruct the missing data within the masked region. In most of the cases, this is however a very difficult task, as the clustering properties of the considered field (e.g. galaxy redshift surveys or 21cm surveys) are not well understood theoretically (see discussion about summary statistics above). On the other hand, the statistical properties of the considered field can be learned by neural networks and used to reconstruct the masked region. This idea has been developed in the machine learning community to \textit{inpaint} the missing pixels of images \citep{Pathak_2016, Yang_2016, Demir_2018,yan2018shift,yu2018generative,liu18,nazeri2019edgeconnect,yu2019free,zhu21}.

The use of image inpainting techniques based on deep learning has recently gained increasing interest in the cosmological community. Several works have successfully used deep convolutional neural networks to reconstruct missing data in 2D maps of the cosmic microwave background \citep{srinivasan19,kai20,alireza21,montefalcone21} and in the galactic foreground intensity and polarization maps \citep{giuseppe20}.

In this work we use these techniques to investigate whether we can reconstruct masked regions from 2D images generated from state-of-the-art magneto-hydrodynamic simulations. For this, we make use of data from the Cosmology and Astrophysics with MachinE Learning Simulations \citep[CAMELS,][]{CAMELS} Multifield Dataset \citep[CMD,][]{CMD}, a collection of hundreds of thousands of 2D maps and 3D grids containing 13 different fields from thousands of different cosmological and astrophysical models. To our knowledge, this is the first time that such a study is carried out with data from state-of-the-art hydrodynamic simulations over a vast range of cosmological and astrophysical models.

This paper is organised as follows. In Section \ref{sec:Data} we describe the data we use in this work. We outline the architecture and training procedure in Sec. \ref{sec:technique}. The main results of this work are shown in Sec. \ref{sec:results}. Finally, we summarise and discuss the findings of this paper in Sec. \ref{sec:Conclusions}.

%%%%%%%%%%%%%%%%%%%%%%%%%%%%%%%%%%%%%%%%%%%%
\section{Data} \label{sec:Data}								%%%%%%%%%%
%%%%%%%%%%%%%%%%%%%%%%%%%%%%%%%%%%%%%%%%%%%%

In this work we make use of 2D maps from the CAMELS Multifield Dataset\footnote{\url{https://camels-multifield-dataset.readthedocs.io}}, CMD, a collection of hundreds of thousands of 2D maps showing different properties of the gas, dark matter, and stars at $z=0$ from 2,000 state-of-the-art (magneto-)hydrodynamic simulations of the CAMELS project \citep{CAMELS}. All simulations follow the evolution of $256^3$ dark matter particles and $256^3$ fluid elements from $z=127$ down to $z=0$ in a periodic comoving volume of $(25~h^{-1}{\rm Mpc})^3$. Half of the hydrodynamic simulations have been run with the \textsc{AREPO} code \citep{Arepo_public} and employ the same subgrid model as the IllustrisTNG simulations \citep{PillepichA_16a, WeinbergerR_16a}, while the other half has been run with the \textsc{GIZMO} code \citep{Hopkins2015_Gizmo} and utilize the subgrid model of the \textsc{SIMBA} simulation \citep{SIMBA}.

All simulations share the value of these cosmological parameters: baryon density $\Omega_{\rm b}=0.049$, Hubble parameter $h=0.67$, spectral index $n_s=0.96$, sum of the neutrino mass $\sum m_\nu=0$ eV, and the equation-of-state parameter for dark energy $w=-1$. On the other hand, each simulation has a different value of the total matter density parameter $\Omega_{\rm m}$ and $\sigma_8$, the amplitude of the linear power spectrum on scales of $8~h^{-1}{\rm Mpc}$ and also differ in the value of four astrophysical parameters that characterize the efficiency of supernova and Active Galactic Nuclei (AGN) feedback. CMD contains maps for 13 different fields: 1) gas density, 2) gas velocity, 3) gas temperature, 4) gas pressure, 5) gas metallicity, 6) neutral hydrogen density, 7) electron number density, 8) magnetic fields, 9) magnesium-to-iron ratio, 10) dark matter density, 11) dark matter velocity, 12) stellar mass density, 13) total matter density. Each 2D map covers an area of $25\times25~(h^{-1}{\rm Mpc})^2$, contains $256\times256$ pixels, and has a specific value of the cosmological and astrophysical parameters. For each field CMD provides 15,000 maps. We refer the reader to \citep{CMD} for further details on CMD.

In this work we focus our attention on the gas temperature maps, which represent the mass-weighted temperature field of the gas particles in the different simulations.

%%%%%%%%%%%%%%%%%%%%%%%%%%%%%%%%%%%%%%%%%%%%
\section{Technique} \label{sec:technique}						%%%%%%%%%%
%%%%%%%%%%%%%%%%%%%%%%%%%%%%%%%%%%%%%%%%%%%%

In this section we describe the method used to evaluate the performance of the inpainting model. We start describing the construction of the binary masks in Sec. \ref{sec:masks} that we later apply to the CMD maps to mimic the missing data. In Sec. \ref{sec:architecture} we present the architecture of the deep convolutional neural network used to inpaint the masked regions in the data. In Sec. \ref{sec:loss} we discuss the loss function used to train the neural network while the training process is described in Sec. \ref{sec:training}.

\subsection{Binary Masks} \label{sec:masks}
We generate two types of masks: 1) \textit{regular masks} that have either a rectangular or circular shape and cover a continuous portion of the field of view; and 2) \textit{irregular masks} that consist of a set of segments of different width and length randomly placed over the field. For each of these two types, we build masks that cover different fractions of the total area. In particular, we use masks, both regular and irregular, that cover $15\%$ and $30\%$ of the total area. These are realistic numbers that one may encounter in galaxy redshift surveys. In particular, in the Dark Energy Survey (DES) photometric sample for cosmology \citep{des_y3} the masked regions amount to roughly $10\%$ of the total survey area. In spectroscopic surveys such as the Baryon Oscillation Spectroscopic Survey (BOSS) LOWZ and CMASS \citep{BOSS} samples $\sim7\%$ of the total area is lost due to the veto masks. In the extended Baryon Oscillation Spectroscopic Survey (eBOSS) $\sim17\%$ of the area covered by the Luminous Red Galaxy (LRG) and quasar (QSO) \citep{ross20} catalogues was obscured by different types of veto masks. The choice for the sizes of regular masks is straightforward given the desired fraction of the area to be masked. Each irregular mask, on the other hand, is built by successively adding segments of randomly chosen width and length until the number of pixels they cover is the target fraction of the total area. In this paper we will refer to the pixels covered by the mask as the \textit{`hole pixels'} and to the un-masked pixels as \textit{`valid pixels'}.

\subsection{Architecture} \label{sec:architecture}

We use the network architecture presented in \cite{zhu21} based on the so called `Mask-Aware Dynamic Filtering' (MADF) module. This is a deep convolutional neural network consisting of three main stages: the encoder, the recovery decoder and the refinement decoder. The architecture is similar in nature to a U-shaped encoder-decoder network that encodes the semantic information from the valid pixels of the masked image into multiple level feature maps which are later decoded into the low-level pixel values.

The encoder provides the high-level feature maps using the information from the input damaged image and the corresponding binary mask. In particular, rather than using fixed kernels it uses the Mask-Aware Dynamic Filtering (MADF) module to dynamically generate kernels for each convolutional window based on the features of the corresponding position on the mask. The decoder step is further divided into two stages. The recovery decoder performs a rough filling of the holes in the feature maps and produces the first output. A set of refinement decoders are run in parallel to the recovery decoder to refine the decoded feature maps. Another distinct feature of this novel network architecture is the use of the so called `Point-wise Normalisation' (PN) in place of the typical `Batch Normalisation' (BN) in the refinement decoding steps to avoid the `covariant shift' problem arising from the difference between the statistical properties of the features of the hole and valid pixels. We refer the reader to \cite{zhu21} for a detailed discussion of the advantages of this approach.

Although the architecture proposed in \cite{zhu21} is flexible in terms of the model complexity, tuning its hyperparameters would require many tests that are computationally expensive and time demanding. We thus use the same setup proposed in \cite{zhu21} that resulted in excellent results on the benchmark datasets typically used to assess the performance of the image inpainting models. In particular, each of the encoder, recovery decoder and refinement decoders consists of 7 levels with the kernel size and strides of each convolutional operation set empirically. Also the number of refinement decoders is set to be 2 as a compromise between model performance and efficiency.

\subsection{Loss Function} \label{sec:loss}

We use the `inpainting loss' adopted in \cite{liu18} and \cite{zhu21} as the optimisation objective. The total loss function consists of multiple terms that depend on the output of each decoder and are incrementally added. Different loss terms compare different properties of the predicted and the true maps (ground truth).

The first-order comparison is performed using the so called `per-pixel reconstruction loss' that is split into two terms, one evaluated over the valid pixels ($\lvalid$) and one over the hole pixels ($\lhole$),
    \begin{equation}
        \lvalid = \frac{1}{N_{\igt}}\norm{M\odot\(\iout-\igt\)}_1, \label{eq:lvalid}
    \end{equation}
    \begin{equation}
        \lhole = \frac{1}{N_{\igt}}\norm{(1-M)\odot\(\iout-\igt\)}_1. \label{eq:lhole}
    \end{equation}
In Eqs.~\eqref{eq:lvalid} and \eqref{eq:lhole}, $N_{\igt}$ indicates the number of elements in the ground truth map, $M$ is the binary mask, $\iout$ is the model output, $\igt$ is the ground truth image and $\odot$ denotes the elementwise product.

The perceptual loss $\lperc$, introduced by \cite{gatys15}, forces the network to output semantically meaningful predictions as encoded by the feature maps $\Psi_p$ extracted using the \textit{pool1}, \textit{pool2} and \textit{pool3} layers of the pretrained VGG16 ImageNet \citep{simonyan14},
    \begin{equation}
        \lperc = \sum_{p=1}^{3}\frac{\norm{\Psi_p^{\iout}-\Psi_p^{\igt}}_1}{N_{\Psi_p^{\igt}}} + \sum_{p=1}^3\frac{\norm{\Psi_p^{\icomp}-\Psi_p^{\igt}}_1}{N_{\Psi_p^{\igt}}}, \label{eq:lperc}
    \end{equation}
where $N_{\Psi^{\igt}_p}$ denotes the number of elements in the feature map extracted from the VGG16 layer $p$ and $\icomp$ results from the model output with the valid pixels set to their ground truth values.

The style loss $\lstyle$ uses the same feature maps extracted from the VGG16 network as those used for $\lperc$ but computes the L1 loss over their auto-correlation given by the Gram matrix,
    \begin{equation}
        \begin{split}    
        \lstyle =& \sum_{p-1}^3 \frac{\norm{K_{p}\left(\left(\Psi_{p}^{\iout}\right)^{T}\left(\Psi_{p}^{\iout}\right) - \left(\Psi_{p}^{\igt}\right)^{T}\left(\Psi_{p}^{\igt}\right)
        \right)}_1}{C_pC_p} \\
        +&\frac{\norm{K_{p}\left(\left(\Psi_{p}^{\icomp}\right)^{T}\left(\Psi_{p}^{\icomp}\right) - \left(\Psi_{p}^{\igt}\right)^{T}\left(\Psi_{p}^{\igt}\right)
        \right)}_1}{C_pC_p}, \label{eq:lstyle}
        \end{split}
    \end{equation}
where $K_p=1/(C_pH_pW_p)$ is the normalisation factor with $(C_pH_pW_p)$ being the size of the feature vector extracted from layer $p$. The style loss $\lstyle$ helps constraining the texture of the predicted maps to match that of the ground truth.

Finally, the total variation loss $\ltv$ is used to allow for the spatial smoothness in the output map,
    \begin{equation}
        \begin{split}
            \ltv &= \sum_{\left(i,j\right)\in R, \left(i,j+1\right)\in R}\frac{\norm{\icomp^{i,j+1}-\icomp^{i,j}}_1}{N_{\icomp}} \\
            &+ \sum_{\left(i+1,j\right)\in R, \left(i,j\right)\in R}\frac{\norm{\icomp^{i+1,j}-\icomp^{i,j}}_1}{N_{\icomp}}, \label{eq:ltv}
        \end{split}
    \end{equation}
where $R$ represents the 1-pixel dilation of the hole region.

Different loss terms described above are weighted by the corresponding weights and combined to provide the total loss function $\ltot$,
    \begin{equation}
        \ltot = \lvalid + 6\lhole + 0.05\lperc + 120\lstyle + 0.1\ltv. \label{eq:ltot}
    \end{equation}
The weights associated with each term in eq.~\eqref{eq:ltot} are identical to those set by \cite{liu18} found by empirical calibration.

\subsection{Training} \label{sec:training}

In order to train the model we first split the 15,000 IllustrisTNG-based CMD gas temperature maps into the train, validation and test sets. We assign 10,000 maps to the train set, 2,000 to the validation set and 3,000 to the test set.

We train the network using 4 NVIDIA P100 GPUs for 130 epochs. Each epoch consists of multiple iterations with a single iteration using a batch of 16 maps. At a given iteration each map is coupled with a randomly selected binary mask from a pool of 12,000 masks for data augmentation purpose. We apply the $\log_{10}$ transformation to the input maps to reduce the dynamic range of the temperature values and then normalise the training set to zero mean and unit variance using the mean $\mu_{\rm train}$ and standard deviation $\sigma_{\rm train}$ of the train set. The same parameters $\left(\mu_{\rm{train}}, \sigma_{\rm{train}}\right)$ are then used to normalise the $\log_{10}$-transformed validation and test sets. We use the \texttt{Adam} optimizer and set the initial learning rate to 0.0002. We use PyTorch \verb+ReduceLROnPlateau+ function to implement the update policy that decays the learning rate by a factor of 10 if no decrease in the training loss $\ltot$ is observed for 5 consecutive epochs. The training process is completed in approximately 24 hours. After each epoch the model is evaluated on the validation set to monitor any over-fitting to the training set.

%%%%%%%%%%%%%%%%%%%%%%%%%%%%%%%%%%%%%%%%%%%%
\section{Results}\label{sec:results}							%%%%%%%%%%
%%%%%%%%%%%%%%%%%%%%%%%%%%%%%%%%%%%%%%%%%%%%

%%%%%%%%%%%%%%%%%%%%%%%%%%%%%%%%%%%%%%%%%%%%%%%%%%%%%%%%%%%%%%%%%%%%%%%%%%%%%%%%%%%%%%%%
\begin{figure*}
    	\centering
		\includegraphics[width=\paperwidth]{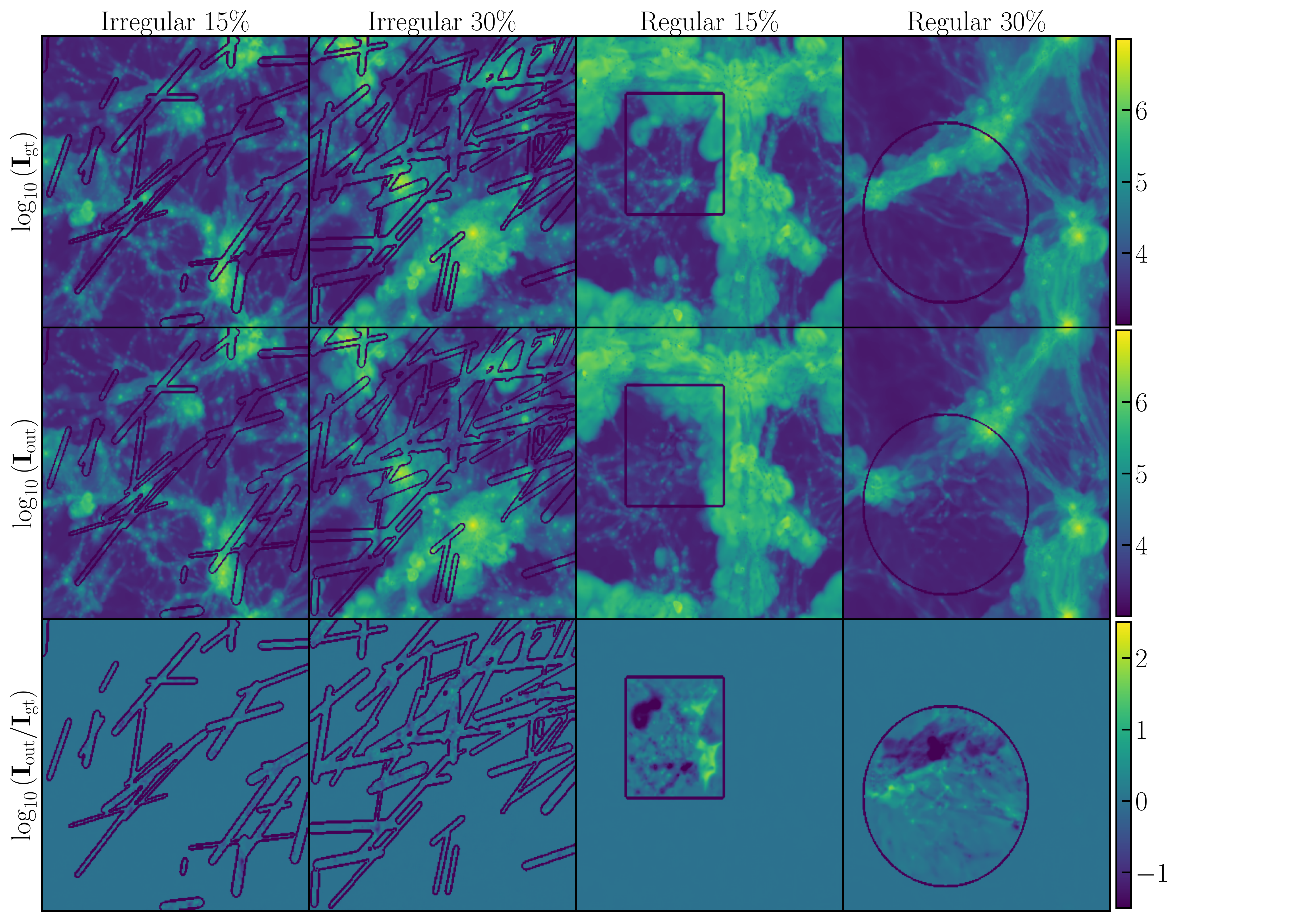}
		\caption{Temperature maps extracted from the CAMELS test set. Different columns display results applying masks of different type and extent. We show the $\log_{10}$ of the temperature maps for better visualisation. Rows in order from top to bottom show the ground truth maps $\log_{10}\igt$, prediction of the network $\log_{10}\iout$ and the difference between the ground truth and the output maps $\log_{10}\left(\iout/\igt\right)$. The contours show the area delimited by the masks. The different panels in the top two rows share the same color-scale while the color range in the bottom row is adapted to highlight the structures in the plot.}  \label{fig:maps}
\end{figure*}
%%%%%%%%%%%%%%%%%%%%%%%%%%%%%%%%%%%%%%%%%%%%%%%%%%%%%%%%%%%%%%%%%%%%%%%%%%%%%%%%%%%%%%%%

We evaluate the model predictions using the holdout test set of 3,000 gas temperature maps and binary masks. None of these maps and masks is exposed to the model during training in order to check how well the results generalise to new data. We first show a visual comparison of the ground truth and predicted maps in Section~\ref{sec:maps_comparison}. We then quantitatively assess the reliability of the inpainted maps using the probability density function in Sec.~\ref{sec:pdf} and the 2D power spectrum in Section~\ref{sec:pk_results}. In Sec.~\ref{sec:auxiliary} we also evaluate the performance of the model in recovering missing data in physical fields different to the one exposed during training.

\subsection{Visual Comparison}\label{sec:maps_comparison}

In Fig.~\ref{fig:maps} we show four temperature maps from the CMD test set. Rather than showing the raw maps we plot the $\log_{10}$ of the temperature values to facilitate a visual inspection. From top to bottom, the first row shows the ground truth maps, the second row displays the output of the reconstruction while the last row shows a pixel-by-pixel comparison between the ground truth and the predicted maps. Different columns show the results for different types (regular or irregular) and extent (fraction of the total area covered) of the binary masks. The left two columns contain results using irregular-shaped masks and a visual comparison between the ground truth and network prediction can barely spot any difference. In the case of regular masks in the right two columns of Fig.~\ref{fig:maps}, there are some clear differences between the target and the predicted map even for the masks with the lower coverage ($15\%$). This naturally arises from the fact that whole structures are wiped-out in the masking process and the inpainting model aims at recovering the correct \textit{style} (or statistical properties) in the reconstructed map rather than matching pixel-by-pixel the output and the ground truth maps. This effect is much more pronounced for the regular masks that cover $30\%$ of the total pixels. Indeed, large structures in the reference map are replaced by an ensemble of smaller structures. This result is not surprising since the lost information cannot be retrieved from the valid pixels given the size of the mask relative to that of the cosmological structures it covers and the size of the whole map. 

We also highlight the near perfect match between the model output and the ground truth maps for the unmasked pixels. This can be attributed to the `skip connection' between the input map and the final stage of the recovery decoder (see Figure~4 in \cite{zhu21}).

Finally, the use of $\ltv$ in the total loss $\ltot$ allows a continuity and smooth transition between the hole and valid pixels. Indeed, in none of the cases tested in this work we find any artefact at the edges of the binary masks.

\subsection{Probability Density Function (PDF)} \label{sec:pdf}

%%%%%%%%%%%%%%%%%%%%%%%%%%%%%%%%%%%%%%%%%%%%%%%%%%%%%%%%%%%%%%%%%%%%%%%%%%%%%%%%%%%%%%%%
\begin{figure}
    	\centering
		\includegraphics[width=\columnwidth]{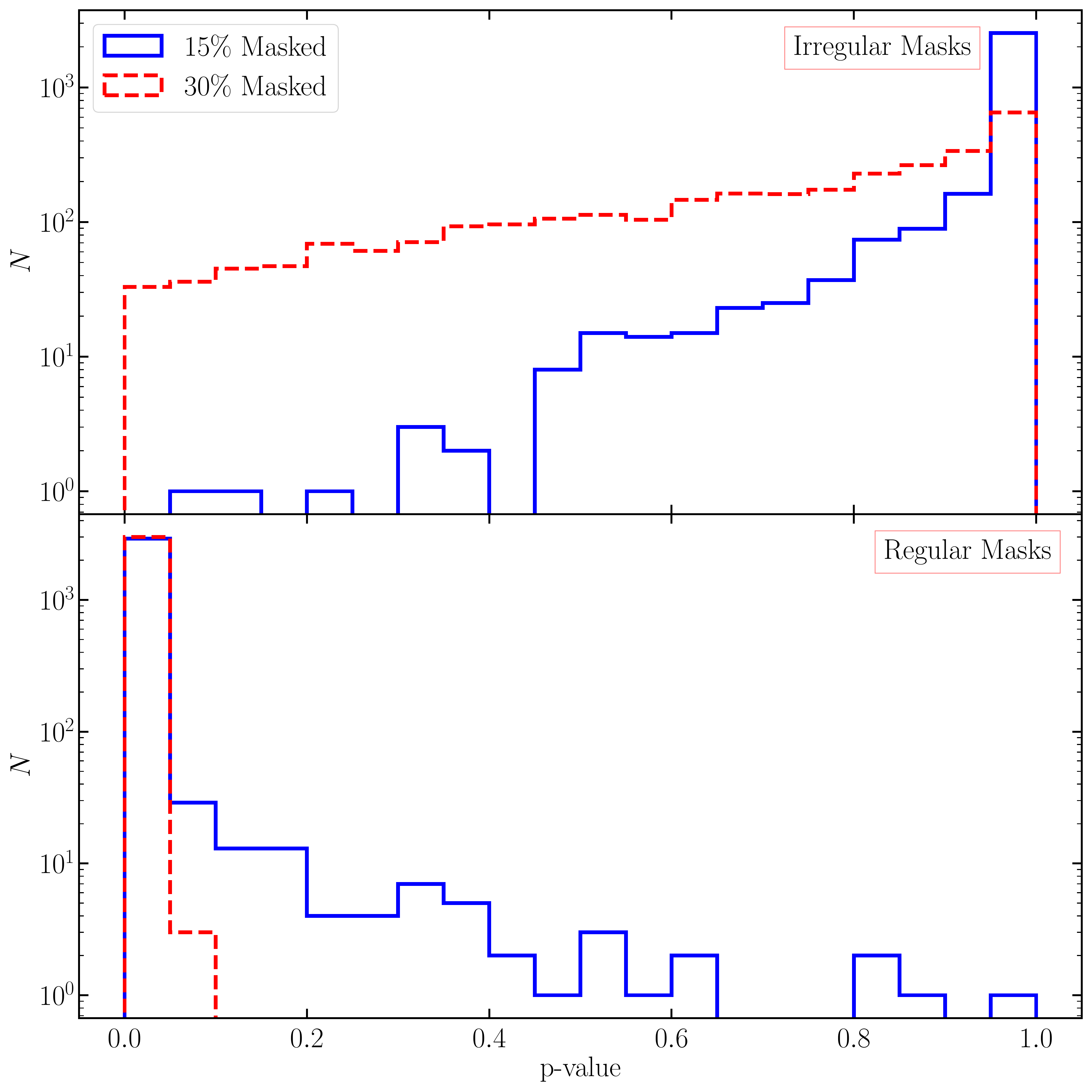}
		\caption{Distribution of the p-values of the Kolmogorov-Smirnov test performed on the 3,000 maps that form the test set. Blue and red histograms show results for masks covering $15\%$ and $30\%$ of the total area, respectively. Top panel: results applying irregular masks. Bottom panel: results when regular-shaped masks are used.}  \label{fig:pdf}
\end{figure}
%%%%%%%%%%%%%%%%%%%%%%%%%%%%%%%%%%%%%%%%%%%%%%%%%%%%%%%%%%%%%%%%%%%%%%%%%%%%%%%%%%%%%%%%

In order to quantify how closely the predicted maps match the ground truth we compare their probability density functions of the temperature values. We use the p-values of the Kolmogorov-Smirnov test (KS-test hereafter) that quantifies the likelihood that the pixels temperature values in the reconstructed and the ground truth maps are drawn from the same underlying distribution. In particular, for each map we estimate the p-value of the KS test by comparing the reconstructed and ground truth map in the masked region and repeat the exercise for all 3,000 maps in the test set. Figure~\ref{fig:pdf} shows the histograms of the corresponding 3,000 p-values for different choices of binary masks.

Under the null hypothesis, i.e. the temperature values in the reconstructed and the ground-truth maps are drawn from the same underlying distribution, we expect a uniform distribution of the KS-test p-values. However, we notice that for irregular masks the distribution peaks at $\text{p-value}=1$ with a near exponential drop at lower p-values indicating an even stronger agreement than that expected between two samples randomly drawn from the same distribution. In order to observe a distribution such as those seen in the top panel of Fig.~\ref{fig:pdf} the temperature values in a non-negligible fraction of the reconstructed pixels must match very closely their counterparts in the ground-truth map. On the other hand, the observed p-values in the case of regular masks in the lower panel of Fig.~\ref{fig:pdf} provide a strong evidence against the null hypothesis especially when $30\%$ of the pixels are masked. Although, for regular masks covering $15\%$ of the data, some of the maps exhibit p-values larger than the threshold of $0.05$, typically used to reject (lower p-values) or accept (larger p-values) the null hypothesis, these form only a relatively small fraction of the 3,000 test maps.

\begin{figure}
    	\centering
		\includegraphics[width=\columnwidth]{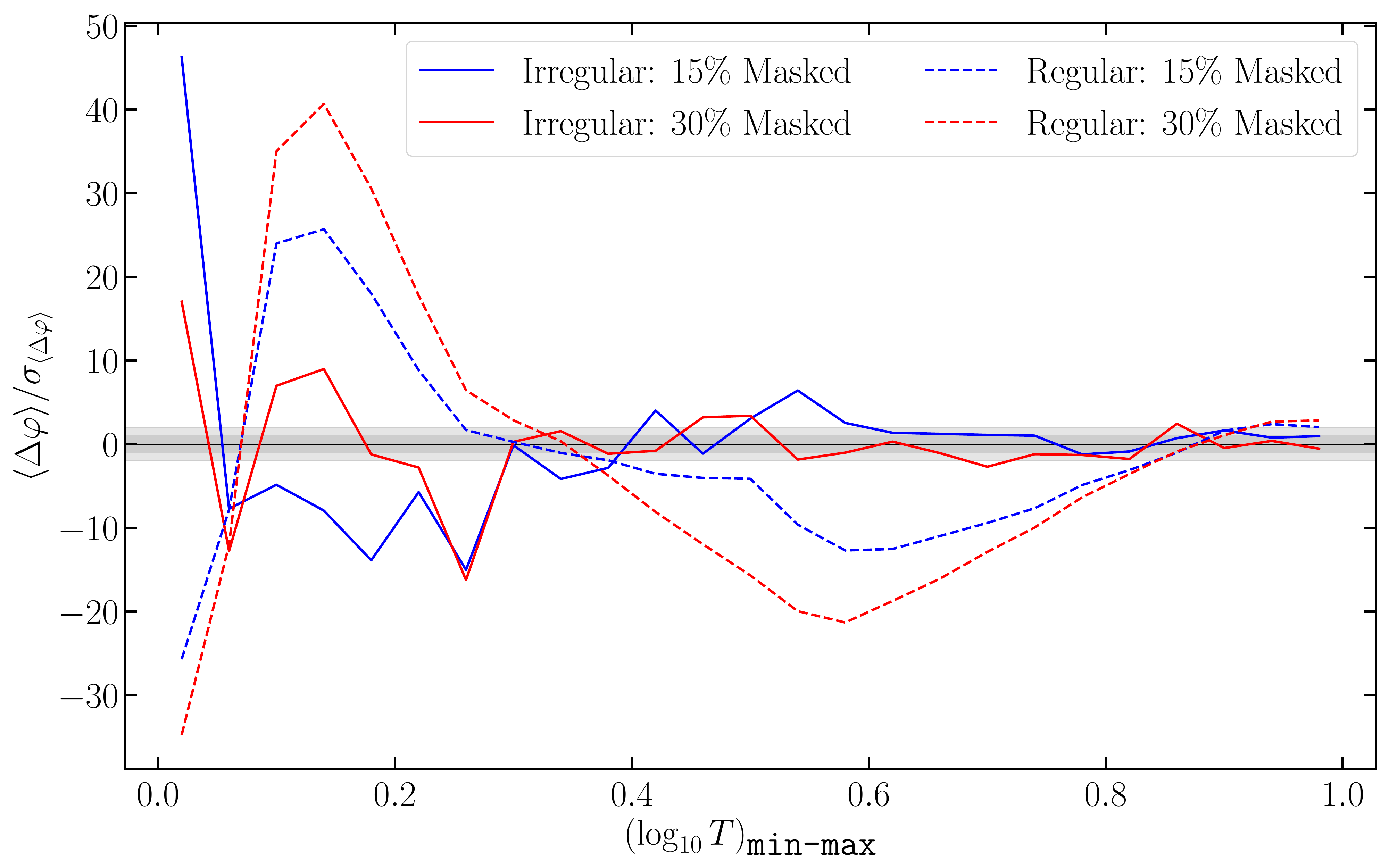}
		\caption{Mean difference between the probability density functions of the \texttt{min-max} scaled and $\log_{10}$-transformed temperature maps, estimated from the model output and the corresponding ground truth map averaged over 3,000 maps in the test set, in units of the standard error on the mean. Continuous lines show results when data are masked using irregular masks while dashed lines correspond to the cases when regular-shaped masks are employed. Blue and red lines correspond to masks covering $15\%$ and $30\%$ of the total area, respectively. Horizontal shaded bands delimit 1-$\sigma$ and 2-$\sigma$ intervals.}  \label{fig:delta_pdf}
\end{figure}

In order to understand the trend observed in Fig.~\ref{fig:pdf} we notice from Fig.~\ref{fig:maps} (last row) that there is a near perfect match in the temperature values between the reconstructed and the ground-truth map near the edges of the mask. However, this is not surprising given the use of the `per-pixel' loss and the `total-variation' loss to train the network that together ensure continuity in the reconstructed map between the hole and valid pixels. On the other hand, the neural network struggles to provide accurate reconstruction in the innermost part of larger masked patches. For irregular masks, a smaller fraction of the area being masked results in a lower probability of different segments that form the mask being joined together to create a single large patch. This increases the fraction of the hole pixels that are close to the mask boundaries where the reconstructed temperature field closely matches its ground-truth values. This explains the blue histogram (irregular masks covering $15\%$ of the data) in the top panel of Fig.~\ref{fig:pdf} being more skewed towards 1 than the red one (irregular masks erasing $30\%$ of the data). For regular masks, along with the aforementioned cause, another effect that contributes to the bad performance seen in Fig.~\ref{fig:pdf} (and later on in Fig.~\ref{fig:delta_pdf} and Fig.~\ref{fig:pk_madf}) is the unique nature of the structures being removed. In particular, the structure that are erased by the regular masks are unique and the network is unable to retrieve the semantic features of the missing data from the valid pixels of the map. The latter effect is field dependent, and we expect a much better performance in terms of recovering both accurate probability density function and the power spectrum for a field that is more homogeneous on the scales of the $(25\mhmpc)^2$ maps, such as the temperature fluctuations seen in the CMB.

%%%%%%%%%%%%%%%%%%%%%%%%%%%%%%%%%%%%%%%%%%%%%%%%%%%%%%%%%%%%%%%%%%%%%%%%%%%%%%%%%%%%%%%%
\begin{table}
\begin{center}
\begin{tabular}{ c | c c c c }
                & Irr. $15\%$  &   Irr. $30\%$    &   Reg. $15\%$  &   Reg. $30\%$  \\
 \hline
 $\Omega_m$     & $-0.041$           &   $+0.055$             &   $-0.040$         &   $-0.036$        \\
 $\sigma_8$     & $+0.075$           &   $+0.084$             &   $-0.028$         &   $-0.029$        \\
 $\rm A_{SN1}$  & $-0.058$           &   $-0.082$             &   $+0.007$         &   $-0.002$        \\
 $\rm A_{AGN1}$ & $+0.001$           &   $+0.007$             &   $-0.010$         &   $+0.013$        \\
 $\rm A_{SN2}$  & $-0.001$           &   $-0.000$             &   $-0.017$         &   $+0.011$        \\
 $\rm A_{AGN2}$ & $+0.007$           &   $+0.022$             &   $-0.007$         &   $-0.041$        \\
 \hline
\end{tabular}
\end{center}
\caption{Pearson correlation coefficient between the values of the 6 simulation parameters and the KS test p-values estimated using the model output for the 3,000 maps from the hold-out test set. The different columns report results for irregular (Irr.) and regular (Reg.) shaped masks covering $15\%$ and $30\%$ of the data. \label{tab:corr_coeff}}
\end{table}
%%%%%%%%%%%%%%%%%%%%%%%%%%%%%%%%%%%%%%%%%%%%%%%%%%%%%%%%%%%%%%%%%%%%%%%%%%%%%%%%%%%%%%%%

We also investigate whether the p-values of the KS test correlate with any of the 6 cosmological or astrophysical parameters used to run the simulations. The Pearson correlation coefficients  reported in Table~\ref{tab:corr_coeff} show that there is no significant correlation between the KS-test p-values and any of the simulation parameters. This indicates that the model performance mainly depends on the properties and extent of the mask and not that much on the particular cosmological and astrophysical model employed.

%%%%%%%%%%%%%%%%%%%%%%%%%%%%%%%%%%%%%%%%%%%%%%%%%%%%%%%%%%%%%%%%%%%%%%%%%%%%%%%%%%%%%%%%
%%%%%%%%%%%%%%%%%%%%%%%%%%%%%%%%%%%%%%%%%%%%%%%%%%%%%%%%%%%%%%%%%%%%%%%%%%%%%%%%%%%%%%%%

While the KS test quantifies the statistical differences between the probability density functions of the ground-truth and the predicted map, it does not indicate where these differences originate from. To investigate if the model's bad performance occurs in specific regimes of the temperature values we compare the corresponding probability density functions estimated from the ground-truth and the predicted maps. In particular, for each map in the test set, we apply the $\log_{10}$ transformation and the \texttt{min-max} scaling to both the ground-truth and the model output before estimating the probability density functions. We show the results in Fig.~\ref{fig:delta_pdf} where the y-axis shows the difference between the two distributions averaged over 3,000 maps from the test set in units of the standard error on the mean as a function of the \texttt{min-max} scaled logarithmic temperature. We note that the disagreement between the model prediction and the ground truth is i) stronger in the low pixel-values regime and improves in pixels with higher intensity of the field; ii) as expected, worse for regular masks compared to the irregular ones and iii) higher for regular masks covering larger extent of the total area.

\subsection{Power Spectrum}\label{sec:pk_results}

%%%%%%%%%%%%%%%%%%%%%%%%%%%%%%%%%%%%%%%%%%%%%%%%%%%%%%%%%%%%%%%%%%%%%%%%%%%%%%%%%%%%%%%%
\begin{figure*}
    	\centering
		\includegraphics[width=\textwidth]{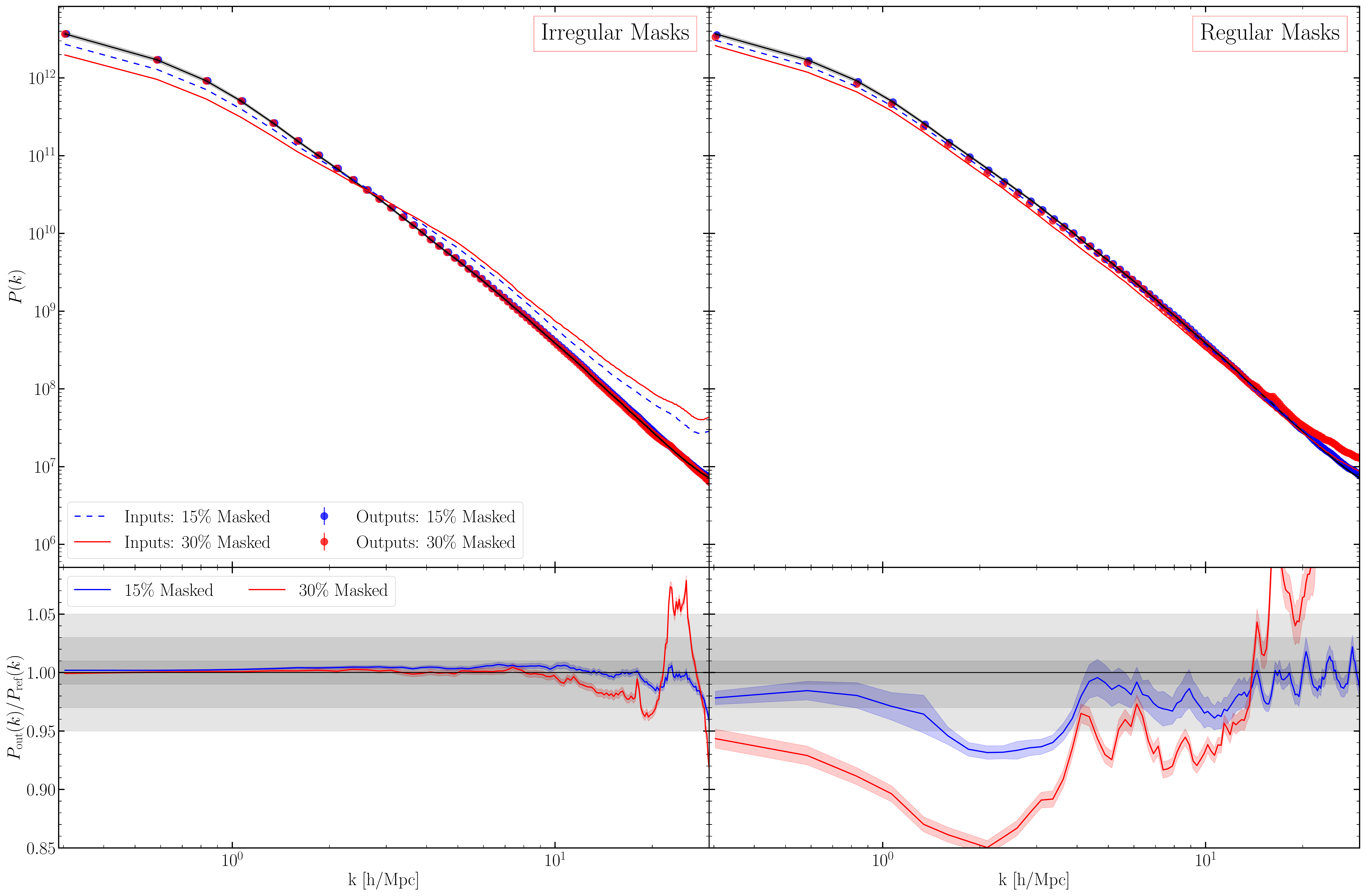}
		\caption{Top panels: Power spectra measured from the ground truth gas temperature maps averaged over 3,000 maps in the test set (black thick line with shaded band), from the input maps masked using irregular-shaped masks (left panels) or regular-shaped masks (right panels) that cover $15\%$ (blue shaded line) and $30\%$ (red thick line) of the total area. Results, after the reconstruction is applied, are shown as blue dots with error bars for masks covering $15\%$ and with red dots with errorbars for those covering $30\%$ of the area. Bottom panels: ratio between the power spectra measured from the reconstructed $P_{\rm out}$ and ground-truth maps $P_{\rm ref}$ are shown when masks cover $15\%$ (blue line with shaded band) and $30\%$ (red line with shaded band) of the area. All errors shown as shaded bands or error bars refer to the error on the mean of 3,000 estimates.}  \label{fig:pk_madf}
\end{figure*}
%%%%%%%%%%%%%%%%%%%%%%%%%%%%%%%%%%%%%%%%%%%%%%%%%%%%%%%%%%%%%%%%%%%%%%%%%%%%%%%%%%%%%%%%

Besides the probability density function, another widely used statistic in cosmology is the power spectrum, defined in this case as,
\begin{equation}
    P(k_1)\delta^D(\vec{k}_1-\vec{k}_2)=\langle F(\vec{k}_1)F(\vec{k}_2)\rangle
\end{equation}
where $F(\vec{k})$ is the Fourier transform of the considered field $F(\vec{x})$, and $\delta^D$ is the Dirac delta. Note that the fields we consider are statistically homogeneous and isotropic, so the power spectrum only depends on the magnitude of the wavenumber, $k$. We use the publicly available \textit{Pylians3}\footnote{\url{https://pylians3.readthedocs.io}} library to compute the power spectra of the maps. In this section we use the power spectrum as a summary statistic to quantify the agreement between the reconstructed maps and their unmasked versions.

The results are shown in Fig.~\ref{fig:pk_madf} for the data from the test set, masked using irregular and regular masks. The top panels show the power spectra measured from the masked data (red thick and blue dashed lines), from the reconstructed maps (blue and red dots with corresponding statistical errors) as well as from the ground truth maps (black thick lines) averaged over the 3,000 maps from the test set. The error bars (on red and blue dots) and shaded bands (around the black thick lines) show the errors on the mean of the 3,000 estimates (i.e. the standard deviation scaled by $\sqrt{3000})$. The differences between the power spectra from the reconstructed and the ground truth maps are barely visible in the top panels. We thus show the ratio between these two quantities in the bottom panels of Fig.~\ref{fig:pk_madf}. As in the top panels, shaded bands in the bottom panels of Fig.~\ref{fig:pk_madf} correspond to the error on the mean of 3,000 estimates. 

For irregular-shaped masks we find that the power spectra of the reconstructed maps agree very well with the reference ones. In particular, for masks covering $15\%$ of the total area the power spectra from the reconstructed maps show a systematic bias with respect to the reference ones of less than $\sim1\%$ up to a wavenumber of $k\sim20~\mihmpc$ (blue dots with errorbars in the top left panel, blue line with shaded band in the bottom left panel of Fig.~\ref{fig:pk_madf}). The accuracy degrades only marginally for wavenumbers below  $k\sim20\rm\mihmpc$, when extending the analysis to irregular masks that cover $30\%$ of the input maps (red points in top left panel and red line in the bottom left panel). For larger wavenumbers (up to the Nyquist wavenumber of $k_{\rm Nyq.}\sim 30\mihmpc$) the power spectra estimated from the reconstructed maps stay accurate within $\sim5\%$ ($\sim10\%$) for irregular masks covering $15\%$ ($30\%$) of the area. 

For regular masks that cover a continuous area of the maps, the reconstruction is less accurate than that for the irregular-shaped masks. As already discussed in Sec.~\ref{sec:pdf}, this is due to the fact that regular-shaped masks erase entire structures in a single large patch. Furthermore, the learning process is also complicated by the fact that we have only very limited number (10) of maps for each set of simulation parameters to train the model, far below the standard size of datasets used to train deep convolutional neural networks. Nevertheless, the network does an excellent job in reconstructing maps where the mask covers $15\%$ of the total area with the recovered power spectra matching the reference ones within $\sim5\%$ up to Nyquist wavenumber of $k_{\rm Nyq.}\sim30\mihmpc$. For regular masks that erase $30\%$ of the data the agreement degrades drastically and becomes strongly scale-dependent.

This analysis, combined with the results in Sec.~\ref{sec:pdf}, shows that the neural network breaks down when large portions of the data are erased in a single patch while results are reliable for the measured power spectrum in other cases explored in this work. One natural way to improve the performance is to train the neural network either on a larger number of simulations for each set of cosmological and astrophysical parameters or on data over an area much larger than the homogeneity scale of the field.

\subsection{Performance on auxiliary data}\label{sec:auxiliary}

%%%%%%%%%%%%%%%%%%%%%%%%%%%%%%%%%%%%%%%%%%%%%%%%%%%%%%%%%%%%%%%%%%%%%%%%%%%%%%%%%%%%%%%%
\begin{figure*}
    	\centering
		\includegraphics[height=0.92\textheight,keepaspectratio]{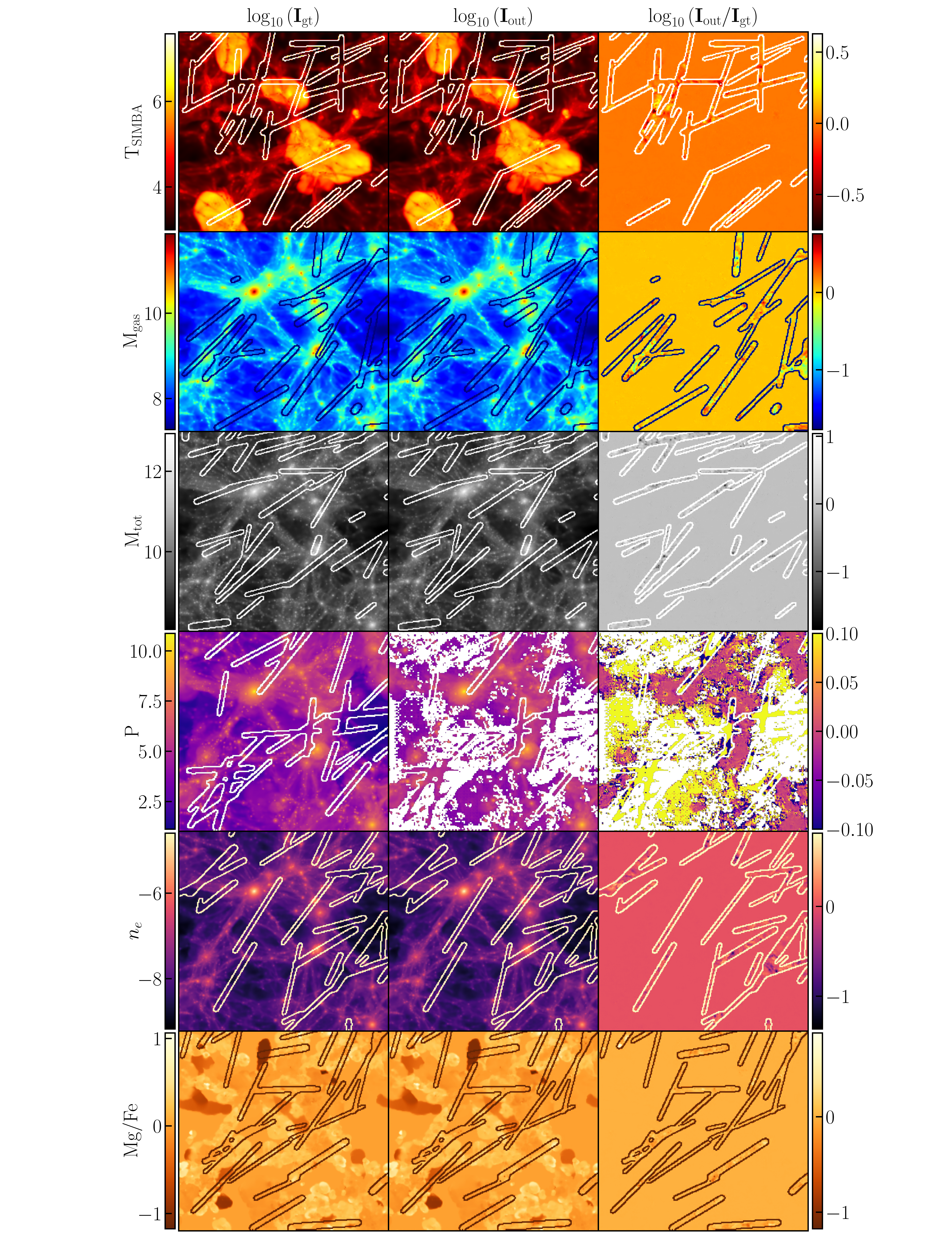}
		\caption{2D maps of 6 different auxiliary fields not used in the network training. From top to bottom row: SIMBA-based gas temperature map ($\rm T_{SIMBA}$), gas density ($\rm M_{gas}$), total matter density ($\rm M_{tot}$), gas pressure ($P$), electron number density ($n_e$), and the magnesium-to-iron ratio (Mg/Fe). Column-wise from left to right: ground truth ($\log_{10}\left(\mathbf{I}_{\rm gt}\right)$), model output ($\log_{10}\left(\mathbf{I}_{\rm out}\right)$) and the difference between the model output and the ground truth maps ($\log_{10}\left(\mathbf{I}_{\rm out}/\mathbf{I}_{\rm gt}\right)$). For a fixed row, the left two columns share the same color coding shown in the color-bars on the left while the the range of the color map in the right-most column is adapted to highlight the differences between the model output and the ground truth map. The color-bars on the right show the color coding for the rightmost column.} \label{fig:maps_auxiliary}
\end{figure*}
%%%%%%%%%%%%%%%%%%%%%%%%%%%%%%%%%%%%%%%%%%%%%%%%%%%%%%%%%%%%%%%%%%%%%%%%%%%%%%%%%%%%%%%%

So far we have used the CMD IllustrisTNG-based gas temperature maps, split into the train, validation and test sets, to both train the model and test its performance on unseen data. In this section we use this model and try to reconstruct the missing data in a number of different fields that are not used during the model training. In particular, we test the performance of the model to recover missing data in maps from other fields such as: 1) SIMBA-based gas temperature maps ($\rm T_{SIMBA}$), 2) gas density ($\rm M_{gas}$), 3) total matter density ($\rm M_{tot}$), 4) gas pressure ($P$), 5) electron density ($n_e$), and 6) the magnesium-to-iron ratio (Mg/Fe). Here we limit the analysis to irregular masks that cover $15\%$ of the total area.

The scales of the pixel intensities in these auxiliary fields are significantly different than the gas temperature field used to train the model. In order to feed the neural network with pixel values that cover a range similar to that of the training set we first rescale each single map to the \texttt{min-max} range of the gas temperature maps in the training set and then normalise it using the $\left(\mu_{\rm{train}}, \sigma_{\rm{train}}\right)$ values used in Sec.~\ref{sec:training}.

The visual comparison between the ground truth and the model output is shown in Fig.~\ref{fig:maps_auxiliary} where each row contains results from a different field. Except for the gas pressure ($P$) maps the differences between the model output and the ground truth are very subtle and can be noticed only through a direct comparison as shown in the rightmost column in Fig.~\ref{fig:maps_auxiliary}. For the gas pressure ($P$) map the model completely fails to recover reliable estimates of the field in specific regions where the model predicts negative pressure. We note that this mainly occurs in the areas with low pixel values in the ground truth maps, in agreement with results shown in Fig.~\ref{fig:delta_pdf}, i.e. the model struggles to provide accurate estimates of the field in the low intensity areas. While this effect is unnoticeable in other fields, it is exacerbated for the gas pressure map. It is very interesting to see that even for fields that have a very different morphology, e.g. Mg/Fe, our model is still able to inpaint features with great success.

%%%%%%%%%%%%%%%%%%%%%%%%%%%%%%%%%%%%%%%%%%%%%%%%%%%%%%%%%%%%%%%%%%%%%%%%%%%%%%%%%%%%%%%%
\begin{figure}
    	\centering
		\includegraphics[width=\columnwidth]{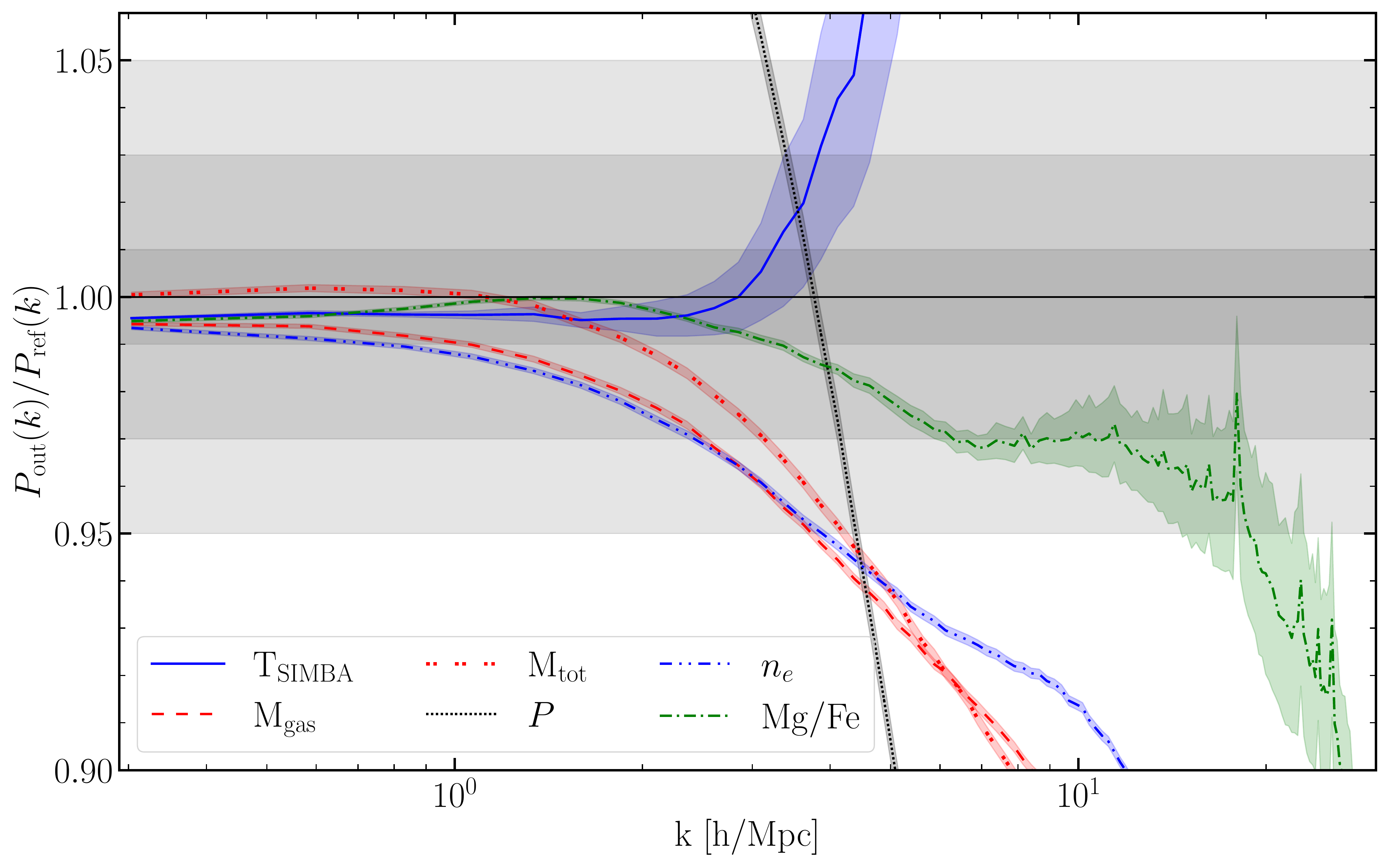}
		\caption{Ratio between the power spectra estimated using the model output and the ground truth averaged over 15,000 maps. Shaded bands show the corresponding errors on the mean. The horizontal bands delimit the $1\%$, $3\%$ and $5\%$ intervals around the reference. Results are shown for 6 different auxiliary fields not used in the network training process.}  \label{fig:pk_auxiliary}
\end{figure}
%%%%%%%%%%%%%%%%%%%%%%%%%%%%%%%%%%%%%%%%%%%%%%%%%%%%%%%%%%%%%%%%%%%%%%%%%%%%%%%%%%%%%%%%

We also perform the analysis using the power spectrum of the auxiliary fields and show the results in Fig.~\ref{fig:pk_auxiliary}. Although results for all 6 fields are worse than those seen in Fig.~\ref{fig:pk_madf} we notice that for the magnesium-iron density field the model is able to match the reference power spectrum within $\sim 5\%$ up to $k\sim 10\mihmpc$. This can be explained by the fact that, as seen in Fig.~\ref{fig:maps_auxiliary}, the structures in the Mg/Fe maps are less complex compared to the other fields. These structures also extend well beyond the typical width of the masks. Interestingly, even for the SIMBA-based gas temperature maps the power spectra of the reconstructed maps are accurate, within $1\%$-level, only for $k<3\mihmpc$ indicating a difference in the small-scale morphological features with respect to the maps based on the IllustrisTNG simulations. The predicted power spectra show a systematic error within $\sim1\%$ up to $k\sim1\mihmpc$ for the gas density ($\rm M_{gas}$) and the electron density ($n_e$) fields. This scale extends to $k\sim2\mihmpc$ for the total matter density ($\rm M_{tot}$) field. On the other hand, the neural network completely fails to return reliable predictions for the gas pressure ($P$) maps resulting in a significantly biased estimates of the power spectra. We do not attempt to provide a physical explanation for these results and leave this for future work.

Our analysis in this section shows that the model does not generalise particularly well to fields it is not exposed to during training. While the model fails to return reliable predictions for some fields, in other cases the validity of the predictions is limited to the largest scales (smallest wavenumbers $k$). These results highlight the need to train a model specifically for the field under investigation. In other words, our model has learned characteristic features of the gas temperature field, that although very generic due to the large variety of cosmological and astrophysical models present in CMD, are still very distinct to those present in other fields.

%%%%%%%%%%%%%%%%%%%%%%%%%%%%%%%%%%%%%%%%%%%%
\section{Summary and Conclusions} \label{sec:Conclusions}		%%%%%%%%%%
%%%%%%%%%%%%%%%%%%%%%%%%%%%%%%%%%%%%%%%%%%%%

In this paper we test the ability of a state-of-the-art deep convolutional neural network architecture, based on the Mask-Aware Dynamic Filtering (MADF) module, to inpaint masked pixels in 2D maps of the CAMELS Multifield Dataset (CMD). We focus our attention on the gas temperature maps based on the IllustrisTNG simulations; CMD provides 15,000 maps obtained from 1,000 state-of-the-art magneto-hydrodynamic simulations with different values of the cosmological and astrophysical parameters.

The dataset is split into a train set of 10,000 maps, a validation set of 2,000 maps and a test set of 3,000 maps. We mimic the missing/masked data in the maps by applying two different kinds of binary masks: 1) regular-shaped that cover a continuous area of each map in a circular or rectangular patch randomly placed within the map and 2) irregular-shaped masks that are composed of a number of segments of various width and length randomly placed across the map area. For each type of mask we test the model performance using two different extents, covering $15\%$ and $30\%$ of the total area. We train the model for 130 epochs using a batch size of 16 for a total of 81,250 training iterations.

We check the model performance using the hold-out test set of 3,000 gas temperature maps and different binary masks. Through a qualitative visual comparison between the model output and the target ground truth, we first show that the model outputs are visually indistinguishable from the ground truth for irregular masks covering either $15\%$ or $30\%$ of the map. The difference becomes more evident for regular-shaped masks. In particular, for regular masks covering $30\%$ of the data in each map, reticular-like artefacts start to appear in correspondence of the masked pixels indicating a breakdown of the model for such a large masks. We also quantify the statistical agreement between the output of the model and the unmasked maps using two different summary statistics: i) the probability density functions and ii) the 2D power spectrum. 

We compare the temperature probability density functions of the model output with that of the ground truth using the Kolmogorov-Smirnov test in the masked regions. We find that, for irregular masks the observed distribution of the KS-test p-values supports the null hypothesis that the reconstructed maps follow the same distribution of the corresponding ground-truth maps. For regular masks, on the other hand, the results of the KS-test indicate that the model fails to match the probability density function of the ground-truth temperature maps. In particular, for regular masks covering $15\%$ of the pixels a vast majority of the 3,000 test maps exhibit a p-value$<0.05$ that indicates a rejection of the null hypothesis. For the largest regular masks that occult $30\%$ of the pixels we find that the KS-test p-values are systematically $\lesssim 0.05$ indicating a strong evidence against the hypothesis that the reconstructed field matches the ground truth in distribution. We do not find any correlation between the KS-test p-values and any of the 6 simulation parameters. We also show that the main source of such a disagreement are the low-intensity pixels.

Estimates of the 2D power spectra highlight an excellent agreement  with a systematic error below $1\text{-}2\%$ up to $k\sim 20\mihmpc$ between the model output and the ground truth when data are masked using irregular masks covering up to $30\%$ of the pixels. The accuracy deteriorates significantly when regular masks are employed, although the systematic offset remains within $5\%$ up to the Nyquist wavenumber $k\sim k_{\rm Nyq.}$ when only $15\%$ of the pixels are masked. The model breaks down when regular masks covering $30\%$ of the total area are used.

The main cause of the model breakdown when data are erased in large patches is the unique nature of the structures being removed combined with a smaller number of maps (for each set of cosmological and astrophysical parameters) used to train the network. On one hand the neural network is unable to retrieve the statistical properties of the missing data from the un-masked pixels, on the other it fails to learn the semantic features of the field from the ensemble of the training maps for a fixed set of cosmological and astrophysical parameters. We thus expect an improvement in the model performance by increasing either the size of each map or the number of maps in the training set.

Finally, we use the model that was trained on the CMD gas temperature maps to perform inpainting on CMD maps of different fields like the SIMBA-based gas temperature maps ($\rm T_{SIMBA}$), the total matter density ($\rm M_{tot}$), the gas density ($\rm M_{gas}$), the gas pressure ($P$), the magnesium-to-iron ratio (Mg/Fe), and the electron density ($n_e$). We find that, even when using irregular masks that extend over $15\%$ of the pixels the model performance degrades significantly compared to when it is applied to the same field it is trained on. An even more important result is that the model performance becomes strongly field-dependent, indicating the need to train the model specifically on the field under investigation.

We conclude that the model used in this work is able to recover reliable pixel-values distributions when data are missing in irregular-shaped patches. These results hold for gas temperature maps that span 1,000 different cosmological and astrophysical models and that exhibit very different morphological aspects such as halos, filaments, and voids. The power spectrum of the inpainted maps exhibit an impressive agreement with their unmasked versions: within $1\%$ for $k_{\rm max}=20~\mihmpc$ and within $5\%$ all the way to the Nyquist wavenumber at $k\sim30~\mihmpc$. For regular-shaped masks our model breaks down in recovering reliable probability density function of the field in the masked patches regardless of the extent while it yields power spectrum estimates accurate at $5\%$ only when $15\%$ of the pixels are masked. This can be a consequence of the very large variety of models seen by the networks; we would expect a higher accuracy also for regular masks if the model was trained on a very large number of images with a fixed cosmological and astrophysical model.

The results presented in this paper have important consequences for cosmological surveys, where missing, masked, and damaged data is a very common issue. This paper paves the way to tackle that problem in different way as it is commonly done in the field. However, more work is needed in order to apply this to real data. We plan to pursue this direction in future work.

\section*{ACKNOWLEDGEMENTS}
This work has made use of the Tiger cluster of Princeton University. 
The CAMELS Multifield Dataset (CMD) is publicly available in \url{https://camels-multifield-dataset.readthedocs.io}. 
Details on the CAMELS simulations can be found in \url{https://www.camel-simulations.org}.

DAA was supported in part by NSF grants AST-2009687 and AST-2108944, and by the Flatiron Institute, which is supported by the Simons Foundation.

\vspace{5mm}

\bibliography{references}{}

\begin{thebibliography}{}
\expandafter\ifx\csname natexlab\endcsname\relax\def\natexlab#1{#1}\fi
\providecommand{\url}[1]{\href{#1}{#1}}
\providecommand{\dodoi}[1]{doi:~\href{http://doi.org/#1}{\nolinkurl{#1}}}
\providecommand{\doeprint}[1]{\href{http://ascl.net/#1}{\nolinkurl{http://ascl.net/#1}}}
\providecommand{\doarXiv}[1]{\href{https://arxiv.org/abs/#1}{\nolinkurl{https://arxiv.org/abs/#1}}}

\bibitem[{{Ade} {et~al.}(2019){Ade}, {Aguirre}, {Ahmed}, {Aiola}, {Ali},
  {Alonso}, {Alvarez}, {Arnold}, {Ashton}, {Austermann}, {Awan}, {Baccigalupi},
  {Baildon}, {Barron}, {Battaglia}, {Battye}, {Baxter}, {Bazarko}, {Beall},
  {Bean}, {Beck}, {Beckman}, {Beringue}, {Bianchini}, {Boada}, {Boettger},
  {Bond}, {Borrill}, {Brown}, {Bruno}, {Bryan}, {Calabrese}, {Calafut},
  {Calisse}, {Carron}, {Challinor}, {Chesmore}, {Chinone}, {Chluba}, {Cho},
  {Choi}, {Coppi}, {Cothard}, {Coughlin}, {Crichton}, {Crowley}, {Crowley},
  {Cukierman}, {D'Ewart}, {D{\"u}nner}, {de Haan}, {Devlin}, {Dicker},
  {Didier}, {Dobbs}, {Dober}, {Duell}, {Duff}, {Duivenvoorden}, {Dunkley},
  {Dusatko}, {Errard}, {Fabbian}, {Feeney}, {Ferraro}, {Flux{\`a}}, {Freese},
  {Frisch}, {Frolov}, {Fuller}, {Fuzia}, {Galitzki}, {Gallardo}, {Tomas Galvez
  Ghersi}, {Gao}, {Gawiser}, {Gerbino}, {Gluscevic}, {Goeckner-Wald}, {Golec},
  {Gordon}, {Gralla}, {Green}, {Grigorian}, {Groh}, {Groppi}, {Guan},
  {Gudmundsson}, {Han}, {Hargrave}, {Hasegawa}, {Hasselfield}, {Hattori},
  {Haynes}, {Hazumi}, {He}, {Healy}, {Henderson}, {Hervias-Caimapo}, {Hill},
  {Hill}, {Hilton}, {Hilton}, {Hincks}, {Hinshaw}, {Hlo{\v{z}}ek}, {Ho}, {Ho},
  {Howe}, {Huang}, {Hubmayr}, {Huffenberger}, {Hughes}, {Ijjas}, {Ikape},
  {Irwin}, {Jaffe}, {Jain}, {Jeong}, {Kaneko}, {Karpel}, {Katayama}, {Keating},
  {Kernasovskiy}, {Keskitalo}, {Kisner}, {Kiuchi}, {Klein}, {Knowles},
  {Koopman}, {Kosowsky}, {Krachmalnicoff}, {Kuenstner}, {Kuo}, {Kusaka},
  {Lashner}, {Lee}, {Lee}, {Leon}, {Leung}, {Lewis}, {Li}, {Li}, {Limon},
  {Linder}, {Lopez-Caraballo}, {Louis}, {Lowry}, {Lungu}, {Madhavacheril},
  {Mak}, {Maldonado}, {Mani}, {Mates}, {Matsuda}, {Maurin}, {Mauskopf}, {May},
  {McCallum}, {McKenney}, {McMahon}, {Meerburg}, {Meyers}, {Miller},
  {Mirmelstein}, {Moodley}, {Munchmeyer}, {Munson}, {Naess}, {Nati},
  {Navaroli}, {Newburgh}, {Nguyen}, {Niemack}, {Nishino}, {Orlowski-Scherer},
  {Page}, {Partridge}, {Peloton}, {Perrotta}, {Piccirillo}, {Pisano},
  {Poletti}, {Puddu}, {Puglisi}, {Raum}, {Reichardt}, {Remazeilles},
  {Rephaeli}, {Riechers}, {Rojas}, {Roy}, {Sadeh}, {Sakurai}, {Salatino},
  {Sathyanarayana Rao}, {Schaan}, {Schmittfull}, {Sehgal}, {Seibert}, {Seljak},
  {Sherwin}, {Shimon}, {Sierra}, {Sievers}, {Sikhosana}, {Silva-Feaver},
  {Simon}, {Sinclair}, {Siritanasak}, {Smith}, {Smith}, {Spergel}, {Staggs},
  {Stein}, {Stevens}, {Stompor}, {Suzuki}, {Tajima}, {Takakura}, {Teply},
  {Thomas}, {Thorne}, {Thornton}, {Trac}, {Tsai}, {Tucker}, {Ullom},
  {Vagnozzi}, {van Engelen}, {Van Lanen}, {Van Winkle}, {Vavagiakis},
  {Verg{\`e}s}, {Vissers}, {Wagoner}, {Walker}, {Ward}, {Westbrook},
  {Whitehorn}, {Williams}, {Williams}, {Wollack}, {Xu}, {Yu}, {Yu}, {Zago},
  {Zhang}, {Zhu}, \& {Simons Observatory Collaboration}}]{simons_obs}
{Ade}, P., {Aguirre}, J., {Ahmed}, Z., {et~al.} 2019, \jcap, 2019, 056,
  \dodoi{10.1088/1475-7516/2019/02/056}

\bibitem[{{Allys} {et~al.}(2020){Allys}, {Marchand}, {Cardoso},
  {Villaescusa-Navarro}, {Ho}, \& {Mallat}}]{Allys_2020}
{Allys}, E., {Marchand}, T., {Cardoso}, J.~F., {et~al.} 2020, \prd, 102,
  103506, \dodoi{10.1103/PhysRevD.102.103506}

\bibitem[{{Banerjee} \& {Abel}(2021{\natexlab{a}})}]{Banerjee_2020}
{Banerjee}, A., \& {Abel}, T. 2021{\natexlab{a}}, \mnras, 500, 5479,
  \dodoi{10.1093/mnras/staa3604}

\bibitem[{{Banerjee} \& {Abel}(2021{\natexlab{b}})}]{Banerjee_2021}
---. 2021{\natexlab{b}}, \mnras, 504, 2911, \dodoi{10.1093/mnras/stab961}

\bibitem[{{Banerjee} {et~al.}(2020){Banerjee}, {Castorina},
  {Villaescusa-Navarro}, {Court}, \& {Viel}}]{Banerjee_2019}
{Banerjee}, A., {Castorina}, E., {Villaescusa-Navarro}, F., {Court}, T., \&
  {Viel}, M. 2020, \jcap, 2020, 032, \dodoi{10.1088/1475-7516/2020/06/032}

\bibitem[{Bayer {et~al.}(2021)Bayer, Villaescusa-Navarro, Massara, Liu,
  Spergel, Verde, Wandelt, Viel, \& Ho}]{Bayer_2021}
Bayer, A.~E., Villaescusa-Navarro, F., Massara, E., {et~al.} 2021, Detecting
  neutrino mass by combining matter clustering, halos, and voids.
\newblock \doarXiv{2102.05049}

\bibitem[{{Bianchi} \& {Verde}(2020)}]{davide20}
{Bianchi}, D., \& {Verde}, L. 2020, \mnras, 495, 1511,
  \dodoi{10.1093/mnras/staa1267}

\bibitem[{{Dai} {et~al.}(2020){Dai}, {Verde}, \& {Xia}}]{Dai_2020}
{Dai}, J.-P., {Verde}, L., \& {Xia}, J.-Q. 2020, \jcap, 2020, 007,
  \dodoi{10.1088/1475-7516/2020/08/007}

\bibitem[{{Dav{\'e}} {et~al.}(2019){Dav{\'e}}, {Angl{\'e}s-Alc{\'a}zar},
  {Narayanan}, {Li}, {Rafieferantsoa}, \& {Appleby}}]{SIMBA}
{Dav{\'e}}, R., {Angl{\'e}s-Alc{\'a}zar}, D., {Narayanan}, D., {et~al.} 2019,
  \mnras, 486, 2827, \dodoi{10.1093/mnras/stz937}

\bibitem[{{Dawson} {et~al.}(2013){Dawson}, {Schlegel}, {Ahn}, {Anderson},
  {Aubourg}, {Bailey}, {Barkhouser}, {Bautista}, {Beifiori}, {Berlind},
  {Bhardwaj}, {Bizyaev}, {Blake}, {Blanton}, {Blomqvist}, {Bolton}, {Borde},
  {Bovy}, {Brandt}, {Brewington}, {Brinkmann}, {Brown}, {Brownstein}, {Bundy},
  {Busca}, {Carithers}, {Carnero}, {Carr}, {Chen}, {Comparat}, {Connolly},
  {Cope}, {Croft}, {Cuesta}, {da Costa}, {Davenport}, {Delubac}, {de Putter},
  {Dhital}, {Ealet}, {Ebelke}, {Eisenstein}, {Escoffier}, {Fan}, {Filiz Ak},
  {Finley}, {Font-Ribera}, {G{\'e}nova-Santos}, {Gunn}, {Guo}, {Haggard},
  {Hall}, {Hamilton}, {Harris}, {Harris}, {Ho}, {Hogg}, {Holder}, {Honscheid},
  {Huehnerhoff}, {Jordan}, {Jordan}, {Kauffmann}, {Kazin}, {Kirkby}, {Klaene},
  {Kneib}, {Le Goff}, {Lee}, {Long}, {Loomis}, {Lundgren}, {Lupton}, {Maia},
  {Makler}, {Malanushenko}, {Malanushenko}, {Mandelbaum}, {Manera}, {Maraston},
  {Margala}, {Masters}, {McBride}, {McDonald}, {McGreer}, {McMahon}, {Mena},
  {Miralda-Escud{\'e}}, {Montero-Dorta}, {Montesano}, {Muna}, {Myers},
  {Naugle}, {Nichol}, {Noterdaeme}, {Nuza}, {Olmstead}, {Oravetz}, {Oravetz},
  {Owen}, {Padmanabhan}, {Palanque-Delabrouille}, {Pan}, {Parejko},
  {P{\^a}ris}, {Percival}, {P{\'e}rez-Fournon}, {P{\'e}rez-R{\`a}fols},
  {Petitjean}, {Pfaffenberger}, {Pforr}, {Pieri}, {Prada}, {Price-Whelan},
  {Raddick}, {Rebolo}, {Rich}, {Richards}, {Rockosi}, {Roe}, {Ross}, {Ross},
  {Rossi}, {Rubi{\~n}o-Martin}, {Samushia}, {S{\'a}nchez}, {Sayres}, {Schmidt},
  {Schneider}, {Sc{\'o}ccola}, {Seo}, {Shelden}, {Sheldon}, {Shen}, {Shu},
  {Slosar}, {Smee}, {Snedden}, {Stauffer}, {Steele}, {Strauss}, {Streblyanska},
  {Suzuki}, {Swanson}, {Tal}, {Tanaka}, {Thomas}, {Tinker}, {Tojeiro},
  {Tremonti}, {Vargas Maga{\~n}a}, {Verde}, {Viel}, {Wake}, {Watson}, {Weaver},
  {Weinberg}, {Weiner}, {West}, {White}, {Wood-Vasey}, {Yeche}, {Zehavi},
  {Zhao}, \& {Zheng}}]{BOSS}
{Dawson}, K.~S., {Schlegel}, D.~J., {Ahn}, C.~P., {et~al.} 2013, \aj, 145, 10,
  \dodoi{10.1088/0004-6256/145/1/10}

\bibitem[{{de la Bella} {et~al.}(2020){de la Bella}, {Tessore}, \&
  {Bridle}}]{Bella_2020}
{de la Bella}, L.~F., {Tessore}, N., \& {Bridle}, S. 2020, arXiv e-prints,
  arXiv:2011.06185.
\newblock \doarXiv{2011.06185}

\bibitem[{{de la Torre} {et~al.}(2013){de la Torre}, {Guzzo}, {Peacock},
  {Branchini}, {Iovino}, {Granett}, {Abbas}, {Adami}, {Arnouts}, {Bel},
  {Bolzonella}, {Bottini}, {Cappi}, {Coupon}, {Cucciati}, {Davidzon}, {De
  Lucia}, {Fritz}, {Franzetti}, {Fumana}, {Garilli}, {Ilbert}, {Krywult}, {Le
  Brun}, {Le F{\`e}vre}, {Maccagni}, {Ma{\l}ek}, {Marulli}, {McCracken},
  {Moscardini}, {Paioro}, {Percival}, {Polletta}, {Pollo}, {Schlagenhaufer},
  {Scodeggio}, {Tasca}, {Tojeiro}, {Vergani}, {Zanichelli}, {Burden}, {Di
  Porto}, {Marchetti}, {Marinoni}, {Mellier}, {Monaco}, {Nichol}, {Phleps},
  {Wolk}, \& {Zamorani}}]{sdlt13}
{de la Torre}, S., {Guzzo}, L., {Peacock}, J.~A., {et~al.} 2013, \aap, 557,
  A54, \dodoi{10.1051/0004-6361/201321463}

\bibitem[{{Demir} \& {Unal}(2018)}]{Demir_2018}
{Demir}, U., \& {Unal}, G. 2018, arXiv e-prints, arXiv:1803.07422.
\newblock \doarXiv{1803.07422}

\bibitem[{{DESI Collaboration} {et~al.}(2016){DESI Collaboration}, {Aghamousa},
  {Aguilar}, {Ahlen}, {Alam}, {Allen}, {Allende Prieto}, {Annis}, {Bailey},
  {Balland}, {Ballester}, {Baltay}, {Beaufore}, {Bebek}, {Beers}, {Bell},
  {Bernal}, {Besuner}, {Beutler}, {Blake}, {Bleuler}, {Blomqvist}, {Blum},
  {Bolton}, {Briceno}, {Brooks}, {Brownstein}, {Buckley-Geer}, {Burden},
  {Burtin}, {Busca}, {Cahn}, {Cai}, {Cardiel-Sas}, {Carlberg}, {Carton},
  {Casas}, {Castand er}, {Cervantes-Cota}, {Claybaugh}, {Close}, {Coker},
  {Cole}, {Comparat}, {Cooper}, {Cousinou}, {Crocce}, {Cuby}, {Cunningham},
  {Davis}, {Dawson}, {de la Macorra}, {De Vicente}, {Delubac}, {Derwent},
  {Dey}, {Dhungana}, {Ding}, {Doel}, {Duan}, {Ealet}, {Edelstein},
  {Eftekharzadeh}, {Eisenstein}, {Elliott}, {Escoffier}, {Evatt}, {Fagrelius},
  {Fan}, {Fanning}, {Farahi}, {Farihi}, {Favole}, {Feng}, {Fernandez},
  {Findlay}, {Finkbeiner}, {Fitzpatrick}, {Flaugher}, {Flender}, {Font-Ribera},
  {Forero-Romero}, {Fosalba}, {Frenk}, {Fumagalli}, {Gaensicke}, {Gallo},
  {Garcia-Bellido}, {Gaztanaga}, {Pietro Gentile Fusillo}, {Gerard},
  {Gershkovich}, {Giannantonio}, {Gillet}, {Gonzalez-de-Rivera},
  {Gonzalez-Perez}, {Gott}, {Graur}, {Gutierrez}, {Guy}, {Habib}, {Heetderks},
  {Heetderks}, {Heitmann}, {Hellwing}, {Herrera}, {Ho}, {Holland}, {Honscheid},
  {Huff}, {Hutchinson}, {Huterer}, {Hwang}, {Illa Laguna}, {Ishikawa},
  {Jacobs}, {Jeffrey}, {Jelinsky}, {Jennings}, {Jiang}, {Jimenez}, {Johnson},
  {Joyce}, {Jullo}, {Juneau}, {Kama}, {Karcher}, {Karkar}, {Kehoe}, {Kennamer},
  {Kent}, {Kilbinger}, {Kim}, {Kirkby}, {Kisner}, {Kitanidis}, {Kneib},
  {Koposov}, {Kovacs}, {Koyama}, {Kremin}, {Kron}, {Kronig}, {Kueter-Young},
  {Lacey}, {Lafever}, {Lahav}, {Lambert}, {Lampton}, {Land riau}, {Lang},
  {Lauer}, {Le Goff}, {Le Guillou}, {Le Van Suu}, {Lee}, {Lee}, {Leitner},
  {Lesser}, {Levi}, {L'Huillier}, {Li}, {Liang}, {Lin}, {Linder}, {Loebman},
  {Luki{\'c}}, {Ma}, {MacCrann}, {Magneville}, {Makarem}, {Manera}, {Manser},
  {Marshall}, {Martini}, {Massey}, {Matheson}, {McCauley}, {McDonald},
  {McGreer}, {Meisner}, {Metcalfe}, {Miller}, {Miquel}, {Moustakas}, {Myers},
  {Naik}, {Newman}, {Nichol}, {Nicola}, {Nicolati da Costa}, {Nie}, {Niz},
  {Norberg}, {Nord}, {Norman}, {Nugent}, {O'Brien}, {Oh}, {Olsen}, {Padilla},
  {Padmanabhan}, {Padmanabhan}, {Palanque-Delabrouille}, {Palmese},
  {Pappalardo}, {P{\^a}ris}, {Park}, {Patej}, {Peacock}, {Peiris}, {Peng},
  {Percival}, {Perruchot}, {Pieri}, {Pogge}, {Pollack}, {Poppett}, {Prada},
  {Prakash}, {Probst}, {Rabinowitz}, {Raichoor}, {Ree}, {Refregier}, {Regal},
  {Reid}, {Reil}, {Rezaie}, {Rockosi}, {Roe}, {Ronayette}, {Roodman}, {Ross},
  {Ross}, {Rossi}, {Rozo}, {Ruhlmann-Kleider}, {Rykoff}, {Sabiu}, {Samushia},
  {Sanchez}, {Sanchez}, {Schlegel}, {Schneider}, {Schubnell}, {Secroun},
  {Seljak}, {Seo}, {Serrano}, {Shafieloo}, {Shan}, {Sharples}, {Sholl},
  {Shourt}, {Silber}, {Silva}, {Sirk}, {Slosar}, {Smith}, {Smoot}, {Som},
  {Song}, {Sprayberry}, {Staten}, {Stefanik}, {Tarle}, {Sien Tie}, {Tinker},
  {Tojeiro}, {Valdes}, {Valenzuela}, {Valluri}, {Vargas-Magana}, {Verde},
  {Walker}, {Wang}, {Wang}, {Weaver}, {Weaverdyck}, {Wechsler}, {Weinberg},
  {White}, {Yang}, {Yeche}, {Zhang}, {Zhao}, {Zheng}, {Zhou}, {Zhou}, {Zhu},
  {Zou}, \& {Zu}}]{DESI_2016}
{DESI Collaboration}, {Aghamousa}, A., {Aguilar}, J., {et~al.} 2016, arXiv
  e-prints, arXiv:1611.00036.
\newblock \doarXiv{1611.00036}

\bibitem[{{Fluri} {et~al.}(2019){Fluri}, {Kacprzak}, {Lucchi}, {Refregier},
  {Amara}, {Hofmann}, \& {Schneider}}]{Fluri_19}
{Fluri}, J., {Kacprzak}, T., {Lucchi}, A., {et~al.} 2019, \prd, 100, 063514,
  \dodoi{10.1103/PhysRevD.100.063514}

\bibitem[{{Friedrich} {et~al.}(2020){Friedrich}, {Uhlemann},
  {Villaescusa-Navarro}, {Baldauf}, {Manera}, \& {Nishimichi}}]{Friedrich_2020}
{Friedrich}, O., {Uhlemann}, C., {Villaescusa-Navarro}, F., {et~al.} 2020,
  \mnras, 498, 464, \dodoi{10.1093/mnras/staa2160}

\bibitem[{{Gatys} {et~al.}(2015){Gatys}, {Ecker}, \& {Bethge}}]{gatys15}
{Gatys}, L.~A., {Ecker}, A.~S., \& {Bethge}, M. 2015, arXiv e-prints,
  arXiv:1508.06576.
\newblock \doarXiv{1508.06576}

\bibitem[{{Giri} \& {Smith}(2020)}]{Giri_2020}
{Giri}, U., \& {Smith}, K.~M. 2020, arXiv e-prints, arXiv:2010.07193.
\newblock \doarXiv{2010.07193}

\bibitem[{{Gualdi} {et~al.}(2021{\natexlab{a}}){Gualdi}, {Gil-Marin}, \&
  {Verde}}]{Gualdi_2021}
{Gualdi}, D., {Gil-Marin}, H., \& {Verde}, L. 2021{\natexlab{a}}, arXiv
  e-prints, arXiv:2104.03976.
\newblock \doarXiv{2104.03976}

\bibitem[{{Gualdi} {et~al.}(2021{\natexlab{b}}){Gualdi}, {Novell},
  {Gil-Mar{\'\i}n}, \& {Verde}}]{Gualdi_2020}
{Gualdi}, D., {Novell}, S., {Gil-Mar{\'\i}n}, H., \& {Verde}, L.
  2021{\natexlab{b}}, \jcap, 2021, 015, \dodoi{10.1088/1475-7516/2021/01/015}

\bibitem[{Gupta {et~al.}(2018)Gupta, Matilla, Hsu, \& Haiman}]{Gupta_18}
Gupta, A., Matilla, J. M.~Z., Hsu, D., \& Haiman, Z. 2018, Phys. Rev. D, 97,
  103515, \dodoi{10.1103/PhysRevD.97.103515}

\bibitem[{{Hahn} \& {Villaescusa-Navarro}(2021)}]{Changhoon_2020}
{Hahn}, C., \& {Villaescusa-Navarro}, F. 2021, \jcap, 2021, 029,
  \dodoi{10.1088/1475-7516/2021/04/029}

\bibitem[{{Hahn} {et~al.}(2020){Hahn}, {Villaescusa-Navarro}, {Castorina}, \&
  {Scoccimarro}}]{Changhoon_2019}
{Hahn}, C., {Villaescusa-Navarro}, F., {Castorina}, E., \& {Scoccimarro}, R.
  2020, \jcap, 2020, 040, \dodoi{10.1088/1475-7516/2020/03/040}

\bibitem[{{Hassan} {et~al.}(2020){Hassan}, {Andrianomena}, \&
  {Doughty}}]{Sultan_2019}
{Hassan}, S., {Andrianomena}, S., \& {Doughty}, C. 2020, \mnras, 494, 5761,
  \dodoi{10.1093/mnras/staa1151}

\bibitem[{{Hopkins}(2015)}]{Hopkins2015_Gizmo}
{Hopkins}, P.~F. 2015, \mnras, 450, 53, \dodoi{10.1093/mnras/stv195}

\bibitem[{{Jeffrey} {et~al.}(2020){Jeffrey}, {Alsing}, \&
  {Lanusse}}]{Niall_2020}
{Jeffrey}, N., {Alsing}, J., \& {Lanusse}, F. 2020, arXiv e-prints,
  arXiv:2009.08459.
\newblock \doarXiv{2009.08459}

\bibitem[{{Kuruvilla} \& {Aghanim}(2021)}]{Kuruvilla_2021}
{Kuruvilla}, J., \& {Aghanim}, N. 2021, arXiv e-prints, arXiv:2102.06709.
\newblock \doarXiv{2102.06709}

\bibitem[{{Laureijs} {et~al.}(2011){Laureijs}, {Amiaux}, {Arduini},
  {Augu{\`e}res}, {Brinchmann}, {Cole}, {Cropper}, {Dabin}, {Duvet}, \&
  {Ealet}}]{Laureijs_2011}
{Laureijs}, R., {Amiaux}, J., {Arduini}, S., {et~al.} 2011, arXiv e-prints,
  arXiv:1110.3193.
\newblock \doarXiv{1110.3193}

\bibitem[{{Liu} {et~al.}(2018){Liu}, {Reda}, {Shih}, {Wang}, {Tao}, \&
  {Catanzaro}}]{liu18}
{Liu}, G., {Reda}, F.~A., {Shih}, K.~J., {et~al.} 2018, arXiv e-prints,
  arXiv:1804.07723.
\newblock \doarXiv{1804.07723}

\bibitem[{{Massara} {et~al.}(2021){Massara}, {Villaescusa-Navarro}, {Ho},
  {Dalal}, \& {Spergel}}]{Massara_2020}
{Massara}, E., {Villaescusa-Navarro}, F., {Ho}, S., {Dalal}, N., \& {Spergel},
  D.~N. 2021, \prl, 126, 011301, \dodoi{10.1103/PhysRevLett.126.011301}

\bibitem[{{Merloni} {et~al.}(2012){Merloni}, {Predehl}, {Becker},
  {B{\"o}hringer}, {Boller}, {Brunner}, {Brusa}, {Dennerl}, {Freyberg},
  {Friedrich}, {Georgakakis}, {Haberl}, {Hasinger}, {Meidinger}, {Mohr},
  {Nandra}, {Rau}, {Reiprich}, {Robrade}, {Salvato}, {Santangelo}, {Sasaki},
  {Schwope}, {Wilms}, \& {German eROSITA Consortium}}]{erosita}
{Merloni}, A., {Predehl}, P., {Becker}, W., {et~al.} 2012, arXiv e-prints,
  arXiv:1209.3114.
\newblock \doarXiv{1209.3114}

\bibitem[{{Mohammad} {et~al.}(2020){Mohammad}, {Percival}, {Seo}, {Chapman},
  {Bianchi}, {Ross}, {Zhao}, {Lang}, {Bautista}, {Brinkmann}, {Brownstein},
  {Burtin}, {Chuang}, {Dawson}, {de la Torre}, {de Mattia}, {Eftekharzadeh},
  {Fromenteau}, {Gil-Mar{\'\i}n}, {Hou}, {Mueller}, {Neveux}, {Paviot},
  {Raichoor}, {Rossi}, {Schneider}, {Tamone}, {Tinker}, {Tojeiro}, {Vargas
  Maga{\~n}a}, \& {Zhao}}]{faizan20}
{Mohammad}, F.~G., {Percival}, W.~J., {Seo}, H.-J., {et~al.} 2020, \mnras, 498,
  128, \dodoi{10.1093/mnras/staa2344}

\bibitem[{{Montefalcone} {et~al.}(2021){Montefalcone}, {Abitbol}, {Kodwani}, \&
  {Grumitt}}]{montefalcone21}
{Montefalcone}, G., {Abitbol}, M.~H., {Kodwani}, D., \& {Grumitt}, R.~D.~P.
  2021, \jcap, 2021, 055, \dodoi{10.1088/1475-7516/2021/03/055}

\bibitem[{Nazeri {et~al.}(2019)Nazeri, Ng, Joseph, Qureshi, \&
  Ebrahimi}]{nazeri2019edgeconnect}
Nazeri, K., Ng, E., Joseph, T., Qureshi, F.~Z., \& Ebrahimi, M. 2019, arXiv
  preprint arXiv:1901.00212

\bibitem[{{Ntampaka} {et~al.}(2019){Ntampaka}, {Eisenstein}, {Yuan}, \&
  {Garrison}}]{Ntampaka_19}
{Ntampaka}, M., {Eisenstein}, D.~J., {Yuan}, S., \& {Garrison}, L.~H. 2019,
  arXiv e-prints, arXiv:1909.10527.
\newblock \doarXiv{1909.10527}

\bibitem[{{Pathak} {et~al.}(2016){Pathak}, {Krahenbuhl}, {Donahue}, {Darrell},
  \& {Efros}}]{Pathak_2016}
{Pathak}, D., {Krahenbuhl}, P., {Donahue}, J., {Darrell}, T., \& {Efros}, A.~A.
  2016, arXiv e-prints, arXiv:1604.07379.
\newblock \doarXiv{1604.07379}

\bibitem[{{Pillepich} {et~al.}(2018){Pillepich}, {Springel}, {Nelson}, {Genel},
  {Naiman}, {Pakmor}, {Hernquist}, {Torrey}, {Vogelsberger}, {Weinberger}, \&
  {Marinacci}}]{PillepichA_16a}
{Pillepich}, A., {Springel}, V., {Nelson}, D., {et~al.} 2018, \mnras, 473,
  4077, \dodoi{10.1093/mnras/stx2656}

\bibitem[{{Puglisi} \& {Bai}(2020)}]{giuseppe20}
{Puglisi}, G., \& {Bai}, X. 2020, \apj, 905, 143,
  \dodoi{10.3847/1538-4357/abc47c}

\bibitem[{{Raghunathan} {et~al.}(2019){Raghunathan}, {Holder}, {Bartlett},
  {Patil}, {Reichardt}, \& {Whitehorn}}]{srinivasan19}
{Raghunathan}, S., {Holder}, G.~P., {Bartlett}, J.~G., {et~al.} 2019, \jcap,
  2019, 037, \dodoi{10.1088/1475-7516/2019/11/037}

\bibitem[{{Ravanbakhsh} {et~al.}(2017){Ravanbakhsh}, {Oliva}, {Fromenteau},
  {Price}, {Ho}, {Schneider}, \& {Poczos}}]{Siamak_16}
{Ravanbakhsh}, S., {Oliva}, J., {Fromenteau}, S., {et~al.} 2017, arXiv
  e-prints, arXiv:1711.02033.
\newblock \doarXiv{1711.02033}

\bibitem[{{Ribli} {et~al.}(2019){Ribli}, {Pataki}, {Zorrilla Matilla}, {Hsu},
  {Haiman}, \& {Csabai}}]{Ribli_19}
{Ribli}, D., {Pataki}, B.~{\'A}., {Zorrilla Matilla}, J.~M., {et~al.} 2019,
  \mnras, 490, 1843, \dodoi{10.1093/mnras/stz2610}

\bibitem[{{Ross} {et~al.}(2012){Ross}, {Percival}, {S{\'a}nchez}, {Samushia},
  {Ho}, {Kazin}, {Manera}, {Reid}, {White}, {Tojeiro}, {McBride}, {Xu}, {Wake},
  {Strauss}, {Montesano}, {Swanson}, {Bailey}, {Bolton}, {Dorta}, {Eisenstein},
  {Guo}, {Hamilton}, {Nichol}, {Padmanabhan}, {Prada}, {Schlegel},
  {Maga{\~n}a}, {Zehavi}, {Blanton}, {Bizyaev}, {Brewington}, {Cuesta},
  {Malanushenko}, {Malanushenko}, {Oravetz}, {Parejko}, {Pan}, {Schneider},
  {Shelden}, {Simmons}, {Snedden}, \& {Zhao}}]{ross12}
{Ross}, A.~J., {Percival}, W.~J., {S{\'a}nchez}, A.~G., {et~al.} 2012, \mnras,
  424, 564, \dodoi{10.1111/j.1365-2966.2012.21235.x}

\bibitem[{{Ross} {et~al.}(2020){Ross}, {Bautista}, {Tojeiro}, {Alam}, {Bailey},
  {Burtin}, {Comparat}, {Dawson}, {de Mattia}, {du Mas des Bourboux},
  {Gil-Mar{\'\i}n}, {Hou}, {Kong}, {Lyke}, {Mohammad}, {Moustakas}, {Mueller},
  {Myers}, {Percival}, {Raichoor}, {Rezaie}, {Seo}, {Smith}, {Tinker},
  {Zarrouk}, {Zhao}, {Zhao}, {Bizyaev}, {Brinkmann}, {Brownstein}, {Rosell},
  {Chabanier}, {Choi}, {Chuang}, {Cruz-Gonzalez}, {de la Macorra}, {de la
  Torre}, {Escoffier}, {Fromenteau}, {Higley}, {Jullo}, {Kneib}, {McLane},
  {Mu{\~n}oz-Guti{\'e}rrez}, {Neveux}, {Newman}, {Nitschelm},
  {Palanque-Delabrouille}, {Paviot}, {Pullen}, {Rossi}, {Ruhlmann-Kleider},
  {Schneider}, {Maga{\~n}a}, {Vivek}, \& {Zhang}}]{ross20}
{Ross}, A.~J., {Bautista}, J., {Tojeiro}, R., {et~al.} 2020, \mnras, 498, 2354,
  \dodoi{10.1093/mnras/staa2416}

\bibitem[{{Samushia} {et~al.}(2021){Samushia}, {Slepian}, \&
  {Villaescusa-Navarro}}]{Samushia_2021}
{Samushia}, L., {Slepian}, Z., \& {Villaescusa-Navarro}, F. 2021, arXiv
  e-prints, arXiv:2102.01696.
\newblock \doarXiv{2102.01696}

\bibitem[{{Schmelzle} {et~al.}(2017){Schmelzle}, {Lucchi}, {Kacprzak}, {Amara},
  {Sgier}, {R{\'e}fr{\'e}gier}, \& {Hofmann}}]{Schmelzle_17}
{Schmelzle}, J., {Lucchi}, A., {Kacprzak}, T., {et~al.} 2017, arXiv e-prints,
  arXiv:1707.05167.
\newblock \doarXiv{1707.05167}

\bibitem[{{Sevilla-Noarbe} {et~al.}(2021){Sevilla-Noarbe}, {Bechtol}, {Carrasco
  Kind}, {Carnero Rosell}, {Becker}, {Drlica-Wagner}, {Gruendl}, {Rykoff},
  {Sheldon}, {Yanny}, {Alarcon}, {Allam}, {Amon}, {Benoit-L{\'e}vy},
  {Bernstein}, {Bertin}, {Burke}, {Carretero}, {Choi}, {Diehl}, {Everett},
  {Flaugher}, {Gaztanaga}, {Gschwend}, {Harrison}, {Hartley}, {Hoyle},
  {Jarvis}, {Johnson}, {Kessler}, {Kron}, {Kuropatkin}, {Leistedt}, {Li},
  {Menanteau}, {Morganson}, {Ogando}, {Palmese}, {Paz-Chinch{\'o}n}, {Pieres},
  {Pond}, {Rodriguez-Monroy}, {Smith}, {Stringer}, {Troxel}, {Tucker}, {de
  Vicente}, {Wester}, {Zhang}, {Abbott}, {Aguena}, {Annis}, {Avila},
  {Bhargava}, {Bridle}, {Brooks}, {Brout}, {Castander}, {Cawthon}, {Chang},
  {Conselice}, {Costanzi}, {Crocce}, {da Costa}, {Pereira}, {Davis}, {Desai},
  {Dietrich}, {Doel}, {Eckert}, {Evrard}, {Ferrero}, {Fosalba},
  {Garc{\'\i}a-Bellido}, {Gerdes}, {Giannantonio}, {Gruen}, {Gutierrez},
  {Hinton}, {Hollowood}, {Honscheid}, {Huff}, {Huterer}, {James}, {Jeltema},
  {Kuehn}, {Lahav}, {Lidman}, {Lima}, {Lin}, {Maia}, {Marshall}, {Martini},
  {Melchior}, {Miquel}, {Mohr}, {Morgan}, {Neilsen}, {Plazas}, {Romer},
  {Roodman}, {Sanchez}, {Scarpine}, {Schubnell}, {Serrano}, {Smith}, {Suchyta},
  {Tarle}, {Thomas}, {To}, {Varga}, {Wechsler}, {Weller}, {Wilkinson}, \& {DES
  Collaboration}}]{des_y3}
{Sevilla-Noarbe}, I., {Bechtol}, K., {Carrasco Kind}, M., {et~al.} 2021, \apjs,
  254, 24, \dodoi{10.3847/1538-4365/abeb66}

\bibitem[{{Simonyan} \& {Zisserman}(2014)}]{simonyan14}
{Simonyan}, K., \& {Zisserman}, A. 2014, arXiv e-prints, arXiv:1409.1556.
\newblock \doarXiv{1409.1556}

\bibitem[{{Spergel} {et~al.}(2015){Spergel}, {Gehrels}, {Baltay}, {Bennett},
  {Breckinridge}, {Donahue}, {Dressler}, {Gaudi}, {Greene}, {Guyon}, {Hirata},
  {Kalirai}, {Kasdin}, {Macintosh}, {Moos}, {Perlmutter}, {Postman},
  {Rauscher}, {Rhodes}, {Wang}, {Weinberg}, {Benford}, {Hudson}, {Jeong},
  {Mellier}, {Traub}, {Yamada}, {Capak}, {Colbert}, {Masters}, {Penny},
  {Savransky}, {Stern}, {Zimmerman}, {Barry}, {Bartusek}, {Carpenter}, {Cheng},
  {Content}, {Dekens}, {Demers}, {Grady}, {Jackson}, {Kuan}, {Kruk}, {Melton},
  {Nemati}, {Parvin}, {Poberezhskiy}, {Peddie}, {Ruffa}, {Wallace}, {Whipple},
  {Wollack}, \& {Zhao}}]{roman_obs}
{Spergel}, D., {Gehrels}, N., {Baltay}, C., {et~al.} 2015, arXiv e-prints,
  arXiv:1503.03757.
\newblock \doarXiv{1503.03757}

\bibitem[{{Square Kilometre Array Cosmology Science Working Group}
  {et~al.}(2018){Square Kilometre Array Cosmology Science Working Group},
  {Bacon}, {Battye}, {Bull}, {Camera}, {Ferreira}, {Harrison}, {Parkinson},
  {Pourtsidou}, {Santos}, {Wolz}, {Abdalla}, {Akrami}, {Alonso},
  {Andrianomena}, {Ballardini}, {Bernal}, {Bertacca}, {Bengaly}, {Bonaldi},
  {Bonvin}, {Brown}, {Chapman}, {Chen}, {Chen}, {Cunnington}, {Davis},
  {Dickinson}, {Fonseca}, {Grainge}, {Harper}, {Jarvis}, {Maartens}, {Maddox},
  {Padmanabhan}, {Pritchard}, {Raccanelli}, {Rivi}, {Roychowdhury}, {Sahlen},
  {Schwarz}, {Siewert}, {Viel}, {Villaescusa-Navarro}, {Xu}, {Yamauchi}, \&
  {Zuntz}}]{SKA_paper}
{Square Kilometre Array Cosmology Science Working Group}, {Bacon}, D.~J.,
  {Battye}, R.~A., {et~al.} 2018, arXiv e-prints, arXiv:1811.02743.
\newblock \doarXiv{1811.02743}

\bibitem[{{Tamura} {et~al.}(2016){Tamura}, {Takato}, {Shimono}, {Moritani},
  {Yabe}, {Ishizuka}, {Ueda}, {Kamata}, {Aghazarian}, {Arnouts}, {Barban},
  {Barkhouser}, {Borges}, {Braun}, {Carr}, {Chabaud}, {Chang}, {Chen}, {Chiba},
  {Chou}, {Chu}, {Cohen}, {de Almeida}, {de Oliveira}, {de Oliveira}, {Dekany},
  {Dohlen}, {dos Santos}, {dos Santos}, {Ellis}, {Fabricius}, {Ferrand},
  {Ferreira}, {Golebiowski}, {Greene}, {Gross}, {Gunn}, {Hammond}, {Harding},
  {Hart}, {Heckman}, {Hirata}, {Ho}, {Hope}, {Hovland}, {Hsu}, {Hu}, {Huang},
  {Jaquet}, {Jing}, {Karr}, {Kimura}, {King}, {Komatsu}, {Le Brun}, {Le
  F{\`e}vre}, {Le Fur}, {Le Mignant}, {Ling}, {Loomis}, {Lupton}, {Madec},
  {Mao}, {Marrara}, {Mendes de Oliveira}, {Minowa}, {Morantz}, {Murayama},
  {Murray}, {Ohyama}, {Orndorff}, {Pascal}, {Pereira}, {Reiley}, {Reinecke},
  {Ritter}, {Roberts}, {Schwochert}, {Seiffert}, {Smee}, {Sodre}, {Spergel},
  {Steinkraus}, {Strauss}, {Surace}, {Suto}, {Suzuki}, {Swinbank}, {Tait},
  {Takada}, {Tamura}, {Tanaka}, {Tresse}, {Verducci}, {Vibert}, {Vidal},
  {Wang}, {Wen}, {Yan}, \& {Yasuda}}]{pfs2016}
{Tamura}, N., {Takato}, N., {Shimono}, A., {et~al.} 2016, in Society of
  Photo-Optical Instrumentation Engineers (SPIE) Conference Series, Vol. 9908,
  Ground-based and Airborne Instrumentation for Astronomy VI, 99081M,
  \dodoi{10.1117/12.2232103}

\bibitem[{{The LSST Dark Energy Science Collaboration} {et~al.}(2018){The LSST
  Dark Energy Science Collaboration}, {Mandelbaum}, {Eifler}, {Hlo{\v{z}}ek},
  {Collett}, {Gawiser}, {Scolnic}, {Alonso}, {Awan}, {Biswas}, {Blazek},
  {Burchat}, {Chisari}, {Dell'Antonio}, {Digel}, {Frieman}, {Goldstein},
  {Hook}, {Ivezi{\'c}}, {Kahn}, {Kamath}, {Kirkby}, {Kitching}, {Krause},
  {Leget}, {Marshall}, {Meyers}, {Miyatake}, {Newman}, {Nichol}, {Rykoff},
  {Sanchez}, {Slosar}, {Sullivan}, \& {Troxel}}]{LSST}
{The LSST Dark Energy Science Collaboration}, {Mandelbaum}, R., {Eifler}, T.,
  {et~al.} 2018, arXiv e-prints, arXiv:1809.01669.
\newblock \doarXiv{1809.01669}

\bibitem[{{Uhlemann} {et~al.}(2020){Uhlemann}, {Friedrich},
  {Villaescusa-Navarro}, {Banerjee}, \& {Codis}}]{Uhlemann_2020}
{Uhlemann}, C., {Friedrich}, O., {Villaescusa-Navarro}, F., {Banerjee}, A., \&
  {Codis}, S. 2020, \mnras, 495, 4006, \dodoi{10.1093/mnras/staa1155}

\bibitem[{{Vafaei Sadr} \& {Farsian}(2021)}]{alireza21}
{Vafaei Sadr}, A., \& {Farsian}, F. 2021, \jcap, 2021, 012,
  \dodoi{10.1088/1475-7516/2021/03/012}

\bibitem[{{Valogiannis} \& {Dvorkin}(2021)}]{Valgiannis_2021}
{Valogiannis}, G., \& {Dvorkin}, C. 2021, arXiv e-prints, arXiv:2108.07821.
\newblock \doarXiv{2108.07821}

\bibitem[{{Villaescusa-Navarro} \& et~al.(2021)}]{CMD}
{Villaescusa-Navarro}, F., \& et~al. 2021, arXiv e-prints, arXiv:2109.XXXXX.
\newblock \doarXiv{2109.XXXXX}

\bibitem[{{Villaescusa-Navarro} {et~al.}(2020){Villaescusa-Navarro}, {Hahn},
  {Massara}, {Banerjee}, {Delgado}, {Ramanah}, {Charnock}, {Giusarma}, {Li},
  {Allys}, {Brochard}, {Uhlemann}, {Chiang}, {He}, {Pisani}, {Obuljen}, {Feng},
  {Castorina}, {Contardo}, {Kreisch}, {Nicola}, {Alsing}, {Scoccimarro},
  {Verde}, {Viel}, {Ho}, {Mallat}, {Wandelt}, \& {Spergel}}]{Quijote}
{Villaescusa-Navarro}, F., {Hahn}, C., {Massara}, E., {et~al.} 2020, \apjs,
  250, 2, \dodoi{10.3847/1538-4365/ab9d82}

\bibitem[{{Villaescusa-Navarro} {et~al.}(2021){Villaescusa-Navarro},
  {Angl{\'e}s-Alc{\'a}zar}, {Genel}, {Spergel}, {Somerville}, {Dave},
  {Pillepich}, {Hernquist}, {Nelson}, {Torrey}, {Narayanan}, {Li}, {Philcox},
  {La Torre}, {Maria Delgado}, {Ho}, {Hassan}, {Burkhart}, {Wadekar},
  {Battaglia}, {Contardo}, \& {Bryan}}]{CAMELS}
{Villaescusa-Navarro}, F., {Angl{\'e}s-Alc{\'a}zar}, D., {Genel}, S., {et~al.}
  2021, \apj, 915, 71, \dodoi{10.3847/1538-4357/abf7ba}

\bibitem[{{Weinberger} {et~al.}(2019){Weinberger}, {Springel}, \&
  {Pakmor}}]{Arepo_public}
{Weinberger}, R., {Springel}, V., \& {Pakmor}, R. 2019, arXiv e-prints,
  arXiv:1909.04667.
\newblock \doarXiv{1909.04667}

\bibitem[{{Weinberger} {et~al.}(2017){Weinberger}, {Springel}, {Hernquist},
  {Pillepich}, {Marinacci}, {Pakmor}, {Nelson}, {Genel}, {Vogelsberger},
  {Naiman}, \& {Torrey}}]{WeinbergerR_16a}
{Weinberger}, R., {Springel}, V., {Hernquist}, L., {et~al.} 2017, \mnras, 465,
  3291, \dodoi{10.1093/mnras/stw2944}

\bibitem[{Yan {et~al.}(2018)Yan, Li, Li, Zuo, \& Shan}]{yan2018shift}
Yan, Z., Li, X., Li, M., Zuo, W., \& Shan, S. 2018, in Proceedings of the
  European conference on computer vision (ECCV), 1--17

\bibitem[{{Yang} {et~al.}(2016){Yang}, {Lu}, {Lin}, {Shechtman}, {Wang}, \&
  {Li}}]{Yang_2016}
{Yang}, C., {Lu}, X., {Lin}, Z., {et~al.} 2016, arXiv e-prints,
  arXiv:1611.09969.
\newblock \doarXiv{1611.09969}

\bibitem[{{Yi} {et~al.}(2020){Yi}, {Guo}, {Fan}, {Hamann}, \& {Wang}}]{kai20}
{Yi}, K., {Guo}, Y., {Fan}, Y., {Hamann}, J., \& {Wang}, Y.~G. 2020, arXiv
  e-prints, arXiv:2001.11651.
\newblock \doarXiv{2001.11651}

\bibitem[{Yu {et~al.}(2018)Yu, Lin, Yang, Shen, Lu, \&
  Huang}]{yu2018generative}
Yu, J., Lin, Z., Yang, J., {et~al.} 2018, in Proceedings of the IEEE conference
  on computer vision and pattern recognition, 5505--5514

\bibitem[{Yu {et~al.}(2019)Yu, Lin, Yang, Shen, Lu, \& Huang}]{yu2019free}
Yu, J., Lin, Z., Yang, J., {et~al.} 2019, in Proceedings of the IEEE/CVF
  International Conference on Computer Vision, 4471--4480

\bibitem[{{Zhu} {et~al.}(2021){Zhu}, {He}, {Li}, {Li}, {Li}, {Liu}, {Ding}, \&
  {Zhang}}]{zhu21}
{Zhu}, M., {He}, D., {Li}, X., {et~al.} 2021, IEEE Transactions on Image
  Processing, 30, 4855, \dodoi{10.1109/TIP.2021.3076310}

\bibitem[{{Zorrilla Matilla} {et~al.}(2020){Zorrilla Matilla}, {Sharma}, {Hsu},
  \& {Haiman}}]{Jose_2020}
{Zorrilla Matilla}, J.~M., {Sharma}, M., {Hsu}, D., \& {Haiman}, Z. 2020, arXiv
  e-prints, arXiv:2007.06529.
\newblock \doarXiv{2007.06529}

\end{thebibliography}
\bibliographystyle{aasjournal}

\end{document}